\documentclass[prx,letterpaper,nobalancelastpage,twocolumn,nofootinbib]{revtex4-2}

\usepackage{graphicx}
\usepackage{amsmath}
\usepackage{amssymb}
\usepackage[english]{babel}
\usepackage{color}
\usepackage[version=4]{mhchem}
\usepackage{hyperref}
\usepackage{dsfont}
\usepackage{placeins}
\usepackage{mathtools}

\usepackage{caption}

\DeclareCaptionLabelSeparator{bar}{ \textbf{\textbar}~}
\usepackage{enumitem}
\setlist{nolistsep}

\newcommand{\YBCO}{$\mathrm{YBa_{2}Cu_{3}O_{7-\delta}}$}

\newcommand{\Caltech}{Engineering \& Applied Science, California Institute of Technology, Pasadena, CA 91125, USA}

\begin{document}

\title{Evidence for Atomic-Scale Inhomogeneity in Superconducting Cuprate NMR}

\author{Jamil Tahir-Kheli}
\affiliation{\Caltech}

\begin{abstract}
In 1990, the Millis, Monien, and Pines (MMP) model and its improvement, the Zha, Barzykin, and Pines (ZBP) model in 1996, emerged as a realistic explanation of the cuprate NMR. These two models assume a single electronic component, translational symmetry, and that the electrons simultaneously have aspects of localized antiferromagnetic (AF) spins and delocalized Cu $d_{x^2-y^2}$ band states. NMR experiments were routinely fit to these models in the 1990s and early 2000s until they finally failed as NMR experiments developed further. It appears that cuprate theorists have given up on explaining the NMR and the NMR data is forgotten. Here, we assume a two-component model of electrons where the electrons reside in two regions, one metallic with delocalized band states, and the other antiferromagnetic with localized spins. This model breaks translational symmetry. We show that the normal state spin relaxation for the planar Cu, O, and Y atoms in $\mathrm{YBa_2Cu_3O_{7-\delta}}$ and their Knight shifts are explained by this two-component model. The temperature dependence of the Cu spin relaxation rate anisotropy in the superconducting state is also explained qualitatively.
\end{abstract}

\maketitle

Surveying the rubble in the field of cuprate theories, one pile of rubble does not belong with the others, in my opinion. This theory should have succeeded because these theorists did the right thing. Rather than telling the electron what it was doing inside the cuprates, they listened to what the electron was telling them.

Specifically, the scene of destruction I am talking about is the work of David Pines and his collaborators that led to the Millis, Monien, and Pines~\cite{MMP} (MMP) model of 1990 for the cuprate NMR and its improvement in 1996, the Zha, Barzykin, and Pines~\cite{ZBP} (ZBP) model.

These workers understood what was already known to NMR resonators and succinctly summarized by Haase et al~\cite{Avramovska2019} in the following quote.
{\it ``Nuclear spins are powerful quantum
sensors of their local environment, so that the versatile methods of
nuclear magnetic resonance (NMR) can be decisive for theories of
condensed matter systems."}

MMP and ZBP created a phenomenological model that was fit to the NMR experimental data. Rather than making a theoretical model and then showing that it fit experiments, they understood that fitting the NMR first and then figuring out what theoretical model could lead to this fit was an excellent way to make progress.

The basic idea is that the NMR (resonance shifts and spin relaxation) can be derived from the dynamic magnetic spin susceptibility, $\chi(\mathbf{q},\omega)$, and the spin hyperfine couplings of the electrons to the nuclei. This relationship between the NMR, $\chi(\mathbf{q},\omega)$, and the hyperfine couplings was obtained by Moriya~\cite{Moriya1963} in 1963 and is a Kubo relation (fluctuation-dissipation theorem) derived by linear response theory. Here, $\mathbf{q}$ and $\omega$ in $\chi(\mathbf{q},\omega)$ are the wave-vector and the angular frequency, respectively. The $\chi(\mathbf{q},\omega)$ in the Moriya expression includes all many-body effects and is exact. For NMR, only $\chi(\mathbf{q=0},\omega=0)$ is necessary for the Knight shift (resonance frequency shift). The ratio of the imaginary part of $\chi(\mathbf{q},\omega)$ and $\omega$, $\mathrm{Im}\chi(\mathbf{q},\omega)/\omega$, as $\omega\rightarrow 0$ is necessary for the spin relaxation. Thus, if the NMR phenomenology is fit to hyperfine couplings and a $\chi(\mathbf{q},\omega)$, then this magnetic susceptibility function contains important physical information about the electronic structure of cuprates. A derivation of the Moriya expression for cuprates can be found in reference~\cite{PenningtonGinsberg2}.

The conceptualization used by MMP and ZBP to suggest functional forms for $\chi(\mathbf{q},\omega)$ is that antiferromagnetic (AF)
spin fluctuations at the planar Cu sites mixed with some kind of delocalized
metallic electron behavior is the physics of the cuprate electrons.
The idea is
that the delocalized electron character leads to the ``metallic" phenomenology
seen in the spin relaxation rates and Knight shifts, while the AF aspect of
the electrons leads to the differences in the temperature dependence of
the spin relaxation rates of planar Cu, planar O, and the Y nuclei
in \YBCO\ due to the positions of the Cu, O, and Y atoms relative to the AF spins.
This conceptualization of cuprates is still believed to be true. Of course, how this mixture of metallic and AF aspects occurs simultaneously is an open question.

In the early years of cuprates, an analysis by Mila and Rice~\cite{MilaRice} estimated the hyperfine couplings. They were found to be close to the values fitted later by Pines et al.  Ab initio quantum chemistry computations also showed that the hyperfine coupling parameters obtained phenomenologically by Pines et al were close to those obtained by calculations~\cite{Renold2001,Meier2001}. These two results suggested the phenomenological models of Pines et al were on the right track. Their phenomenological form for $\chi(\mathbf{q},\omega)$ appeared to correctly capture the mixture of localized AF spin character and delocalized metallic electron character simultaneously.
Hence, until the late 1990s and early 2000s, the MMP and ZBP models were regularly used by NMR resonators to explain their experiments and to sharpen the adjustable parameters in these models.

Unfortunately, the early optimism that the cuprate NMR experiments were understood and could be used to provide constraints and guidance for understanding the mechanism of cuprates fell apart in the late 1990s and early 2000s. Three papers by Charles Pennington and collaborators laid bare two deep problems with MMP and ZBP~\cite{Pennington-PRL-1999,Pennington-STIPDOR,Nandor1999}.

The first problem arose from the MMP and ZBP estimate of an AF spin correlation length of $\xi\approx 2$ lattice spacings. Since $\xi>1$, it was expected that there would be a reduction in the magnitude of the planar Oxygen to neighboring planar Cu nuclear-nuclear spin coupling, $^{17,63}a$ (the superscripts $17$ and $63$ are the atomic numbers of the O and Cu isotopes, respectively), due to the opposite AF spins on the two neighboring Cu atoms. In Yu et al~\cite{Pennington-PRL-1999} and later in Pennington et al~\cite{Pennington-STIPDOR}, $^{17,63}a$ was found to be too large relative to the planar Cu to neighboring planar Cu nuclear-nuclear spin coupling, $^{63,63}a$, to be consistent with antiferromagnetic spin correlations. This result forced Yu et al to conclude, {\it ``The large value of $^{17,63}a/^{63,63}a$ indicates that the form factor cancellation effects do not occur to any significant degree, and hence the antiferromagnetic correlation length must be quite small. The very different spin-lattice relaxation behaviors of $^{17}O$ and $^{63}Cu$ thus return to their former status as an unresolved mystery."}

The second problem was the observation by Nandor et al~\cite{Nandor1999} that the temperature dependent spin relaxation rate of the Y nucleus in \YBCO\ does not ``track" the temperature dependence of its Knight shift and the temperature dependence of the static spin magnetic susceptibility, $\chi_{\mathrm{spin}}(T)$, between $300-700$ K. In particular, MMP and ZBP predict that the spin relaxation rate of Y divided by temperature, $^{89}(1/T_1 T)$, should drop by the same percentage as $\chi_{\mathrm{spin}}(T)$. In fact, $\chi_{\mathrm{spin}}(T)$ dropped by $\approx 13\%$ while $^{89}(1/T_1 T)$ remained constant to within $\pm 4\%$. This unexpected result led Nandor et al to conclude, {\it ``The temperature independence of $\mathit{^{89}(1/T_1 T)}$ is the most serious anomaly of our work; we have developed no quantitative and essentially no qualitative understanding of this finding."}

The catastrophic failure of the MMP and ZBP models that were built phenomenologically using an exact quantum mechanical expression that connects the dynamic magnetic response of the electrons to the nuclei is an extremely serious problem. The only way out of the dilemma is to conclude that one of the foundational assumptions of the model is wrong.

In fact, Yu et al~\cite{Pennington-PRL-1999} proposed a way out. They stated, {\it ``We suggest that to resolve all these issues it may be necessary to invoke a model having $\mathit{\chi(\mathbf{r},\omega)}$, with important dependence on length scales shorter than the lattice constant.''} Here, $\mathbf{r}$ is a position vector, not a wave-vector. In essence, they argue for greater spatial dependence of the dynamic magnetic susceptibility so that the additional spatial freedom allows greater flexibility in the form of $\chi(\mathbf{r},\omega)$.  
Yu et al are proposing that $\chi(\mathbf{r},\omega)$ for $\mathbf{r}$ vectors smaller than the lattice spacing are relevant. In terms of wave-vectors, their proposal is equivalent to including $\mathbf{q}$ in $\chi(\mathbf{q},\omega)$ with $\mathbf{q}$ outside the first Brillouin zone.

In the last $\sim$15 years, many experiments have found evidence for intrinsic inhomogeneity of the electronic structure of cuprates. These experiments are typically interpreted as evidence for some kind of spin-density wave (SDW) or charge-density wave (CDW). In some cases, the wave-vector of the density wave is commensurate with the lattice and in other cases it is incommensurate. In either case some sort of superstructure that can be described by a $\mathbf{k}$ wave-vector is assumed.

After the early 2000s, there was one major theoretical attempt to understand the spin relaxation data by Uldry and Meier in 2005~\cite{Uldry2005}. They did not try to fit the Knight Shift data. This paper did not mention the first problem above and they did not obtain a linear temperature dependence of $^{89}(1/T_1)$ (sometimes written as $^{89}W$) up to $700$ K, as observed. Otherwise theorists abandoned thinking about cuprate NMR by the early 2000s. However, the consensus in the field was that translational symmetry, or momentum $\mathbf{k}$, was a good quantum number. Any defects or impurities merely created scattering centers for electronic states that were well-approximated by $\mathbf{k}$ vectors. Also, translational symmetry was the thinking when MMP and ZBP were formulated. It is the assumption of translational symmetry that sharply restricts the possible choices for $\chi(\mathbf{q},\omega)$. In light of modern experiments, it is reasonable to investigate if removing the translational symmetry assumption intrinsically built into MMP and ZBP may be a way out of the NMR problem.

In this paper, we assume that cuprates are intrinsically inhomogeneous at the atomic-scale and are comprised of two different regions: an insulating AF region with localized electron spins at its planar Cu sites, and a percolating metallic region with delocalized electronic states with orbital character primarily inside this region. 

A specific nucleus in the CuO$\mathrm{_2}$ plane may reside inside the AF or metallic region. It will have neighboring Cu atoms, some of which are in the AF region and some that are in the metallic region. Its Knight shift is an average of the Knight shifts in the two regions. Each regional Knight shift is due to the static magnetic susceptibility, $\chi_M$ and $\chi_{AF}$, for the metallic and AF regions, respectively. We assume the energy exchange between nuclei of the same isotope is fast relative to electronic energy exchange and thereby leads to a nuclear spin temperature. In this case, the total spin relaxation of a particular nuclear isotope is the average of the spin relaxation of all the nuclei of that isotope in the crystal.

We show here that the normal state NMR can be understood with this model of atomic-scale metallic and AF inhomogeneity. 

Finally, the $\chi(\mathbf{q},\omega)$ of MMP and ZBP is only for the normal state. From the experiments of Barrett et al~\cite{Slichter1993,Martindale1991} and Takigawa et al~\cite{Takigawa1991}, there is a temperature dependence of the anisotropy ratio of the planar Cu spin relaxation rate with the applied magnetic field in the CuO$\mathrm{_2}$ plane (ab-axis) to the spin relaxation rate with the applied magnetic field perpendicular to the CuO$\mathrm{_2}$ plane (c-axis), or $^{63}(1/T_1)_{ab}/^{63}(1/T_1)_c$. This ratio drops rapidly from $\sim 3.7$ just above the superconducting transition temperature, $T_c$, to $\approx 2$, and then rises up to $\approx 5$ as the temperature, $T$, goes to zero.

Since there is no known extrapolation of MMP or ZBP to below $T_c$, Bulut and Scalapino~\cite{Bulut1992} and Thelen, Pines, and Lu~\cite{Thelen1993} modeled $\chi(\mathbf{q},\omega)$ as the standard BCS expression for the magnetic susceptibility as a function of the detailed band structure that is then enhanced by an antiferromagnetic Stoner factor. With this $\chi(\mathbf{q},\omega)$, these authors were able to fit the observed Cu relaxation rate anisotropy as a function of temperature to a d-wave superconducting gap symmetry. Their explanation is sensitive to the fine details of the $\mathrm{YBa_2Cu_3O_7}$ band structure.

Their reason for the intial rapid drop in the relaxation rate anisotropy just below $T_c$ is because the d-wave gap opens near wave-vectors $(\pm\pi,0)$ and $(0,\pm\pi)$ and the planar magnetic field relaxation rate drops faster than the out-of-plane relaxation rate as a superconducting gap increases at these wave-vectors. The $T=0$ ratio is obtained by the more rapid drop in the c-axis relaxation rate relative to the ab-axis relaxation near the diagonal ($k_x=\pm k_y$) k-vectors as the d-wave gap freezes out all scattering except in the vicinity of the zero gap directions.

Unlike references~\cite{Bulut1992} and \cite{Thelen1993}, our model is consistent across $T_c$. It also explains this temperature dependent anisotropy ratio. Just below $T_c$ our relaxation rates are dominated by the AF contribution to the relaxation rates, while near $T=0$, the relaxation rates are dominated by the metallic relaxation rates. The initial drop just below $T_c$ occurs for exactly the same reason as references~\cite{Bulut1992} and \cite{Thelen1993}. However, the rise in the anisotropy as $T\rightarrow 0$ is a generic feature of the orbital character of the metallic electronic states near the diagonals ($k_x=\pm k_y$).

\section{Model for Atomic-Scale Inhomogeneity in Cuprates}

We assume a planar Cu is inside the metallic region or the AF region with probabilities $p_M$ and $p_{AF}$ such that

\begin{equation}
\label{psum}
p_M + p_{AF}=1.
\end{equation}

In prior work, we assumed a four-Cu-site plaquette in the metal region was formed for each planar doped hole~\cite{JTK2017,JTK2013}.
Hence, for dopings up to $\sim 0.19$, this assumption led to $p_M=4x$ where $x$ is the hole doping per planar Cu in the CuO$\mathrm{_2}$ planes. For the purposes of this paper, an easy way to approximately relate $p_M$ to doping $x$ is to assume $p_M=0$ when $x=0$ (a pure AF parent compound) and $p_M=1$ (a pure metal) when the $T_c$ dome ends at $x\approx 0.25$. Our relation $p_M=4x$ linearly interpolates between these two endpoints. Hence, at optimal doping of $x=0.16$, $p_M=0.64$ and $p_{AF}=0.36$. These values for $p_M$ and $p_{AF}$ are used for the remainder of this paper.

\begin{figure}[tbp]
\centering \includegraphics[width=0.9\linewidth]
{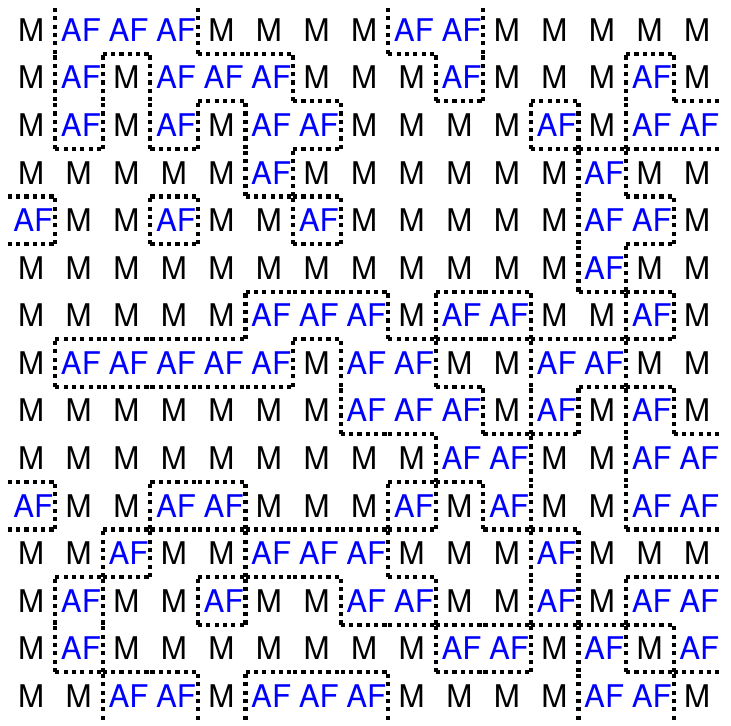}
\caption{Schematic diagram of a randomly doped $15\times 15$ lattice in a single $\mathrm{CuO_2}$ plane with the probability of a Cu site being metallic equal to $p_M=0.64$. The O atoms in the $\mathrm{CuO_2}$ plane are not shown. The black ``M" represents a metallic Cu atom and the blue ``AF" represents an AF Cu atom with a localized spin in the $d_{x^2-y^2}$ orbital. Delocalized electronic band states exist inside the metallic region. The dotted black line shows the boundary between the metallic and AF regions.
}
\label{map}
\end{figure}

The two-dimensional site percolation threshold is $\approx 0.59$. Therefore, the metallic region percolates through the CuO$\mathrm{_2}$ plane and the AF region does not. The AF region is comprised of disconnected islands of antiferromagnetism. The three-dimensional site percolation threshold for a cubic lattice is $\approx 0.31$. Hence, the AF region percolates in 3D. However, the inter-layer AF coupling between unit cells is very small and this coupling normal to the CuO$\mathrm{_2}$ plane direction can be neglected. Hence, the AF regions are essentially 2D. Figure~\ref{map} shows a schematic of a doped $\mathrm{CuO_2}$ plane with $p_M=0.64$.

In this paper, we consider optimal doping only because pseudogap effects are small. Our prior papers proposed that the pseudogap is due to isolated 4-site Cu
plaquettes~\cite{JTK2017,JTK2013}. Here, we do not model the pseudogap.

\section{Model for NMR Spin Relaxation and Knight Shift}

With atomic-scale inhomogeneity, expressing the NMR in terms of a dynamic magnetic spin susceptibility as a function of wave-vector $\mathbf{q}$ and angular frequency $\omega$ is not the best approach. Instead, we will stay in real space. This approach was originally used by Pennington et al~\cite{Pennington1989} in 1989 for the Cu nucleus of \YBCO\ prior to the emergence of MMP and ZBP. After the collapse of these models, Uldry and Meier~\cite{Uldry2005} resurrected this real-space approach in 2005. Neither of these authors considered the possibility of atomic-scale inhomogeneity.

We model the response due to delocalized metallic electrons and localized AF electrons separately. Once the magnetic response in each region is determined, then the overall NMR is calculated for a chosen nucleus by counting the number of its metallic planar Cu and AF planar Cu neighbors.

To model the spin relaxation and Knight shift in cuprates, the temperature dependence of the static magnetic spin susceptibility in the metallic and AF regions, $\chi_{M}(T)$ and $\chi_{AF}(T)$ respectively, and the $T$ dependence of the electron spin correlation times in each region, $\tau_{M}(T)$ and $\tau_{AF}(T)$ respectively are required. These four functions plus the hyperfine couplings in each region and the hyperfine couplings across regions will determine the NMR spin relaxation and Knight shifts.

\subsection{AF Region Spin Correlation Time and Static Magnetic Susceptibility}

Since the AF region is comprised of finite clusters, or islands, of AF spins localized at the planar Cu sites, its spin correlation time, $\tau_{AF}(T)$, should saturate to a constant, $\tau_0$, at high temperatures with a magnitude on the order of $\tau_0\sim\hbar/J_{AF}$ where $J_{AF}$ is the $\mathrm{Cu-Cu}$ AF spin coupling. For \YBCO, $J_{AF}\approx 0.13$ eV $\approx 1500$ K leading to $\hbar/J_{AF}=5.06\times 10^{-15}$ s.

The finite AF clusters will couple to each other through a small spin-spin coupling mediated through the metallic regions between the clusters. This inter-AF-cluster coupling energy scale, $k_B T_x$, should be much smaller than $J_{AF}$, where $k_B$ is Boltzmann's constant and $T_x$ is a temperature. Hence, we expect $k_B T_x<<J_{AF}$.

For $T<<T_x$, $\tau_{AF}(T)$ is due to an electron spin-flip from magnon-magnon scattering because the spin-flip energy is small. If $N(\omega)$ is the magnon density of states and $M(\omega)$ is the energy dependence of the magnon scattering matrix element, then

\begin{equation}
\label{tauaflowT}
\tau_{AF}(T) \sim\int_{0}^{k_B T} N(\omega)^2 M(\omega)^2 d\omega.
\end{equation}

\noindent For 2D AF magnons, $N(\omega)\sim\omega$. Moriya~\cite{Moriya1956,Moriya1956a,Moriya1956b} has shown that $M(\omega)^2$ is the sum of a constant and a $1/\omega^2$ term. At low temperatures, the $1/\omega^2$ term in $M(\omega)^2$ dominates. Hence, $\tau_{AF}(T)\sim T$ for low $T$.

The spin correlation time, $\tau_{AF}$, for the electrons in the AF region is chosen to be of the form,

\begin{equation}
\label{tauaf}
\tau_{AF}(T)=\left(\frac{T}{T + T_x}\right)\tau_0.
\end{equation}

This expression for $\tau_{AF}(T)$ is linear in $T$ at low temperature and saturates to the constant $\tau_0$ at high temperatures.

The Moriya~\cite{Moriya1963} expression applied by MMP and ZBP for the spin relaxation divided by temperature is given by

\begin{eqnarray}
\label{moriyaexp}
\frac{1}{^{k}T_{1\alpha}} & = & \left[\frac{k_B T}{(g\mu_B)^2\hbar^2}\right]
\nonumber \\
& \times &
\left(\frac{1}{N}\right)\sum_{\mathbf{q},\beta\ne\alpha}|^{k}F_{\beta}(\mathbf{q})|^2
\lim_{\omega\rightarrow 0}
\left[\frac{\mathrm{Im}\chi(\mathbf{q},\omega)}{\omega}\right],
\end{eqnarray}

\noindent where $\mu_B$ is the Bohr magneton and $g=2$ is the Land\'{e} g-factor. The superscript $k$ is the atomic number of the nucleus, the subscript $\alpha$ is the magnetic field direction, and $\beta$ is summed over the two directions transverse to the applied magnetic field. $^{k}F_{\beta}(\mathbf{q})$ is the ``form factor"~\cite{Nandor1999} for the atomic number $k$ and direction $\beta$ as a function of the wave-vector $\mathbf{q}$ in the Brillouin zone. $N$ is the number of unit cells.

MMP chose $\lim_{\omega\rightarrow 0}\mathrm{Im}\chi(\mathbf{q},\omega)/\omega\sim \xi^2(T)$ where $\xi(T)$ is the temperature dependence of the AF spin correlation length. They postulated that $\xi(T)$ is of the form

\begin{equation}
\label{xieqn}
\left[\frac{\xi(T)}{\xi(T=0)}\right]^2=\frac{T_x}{T+T_x},
\end{equation}

\noindent for $T>T_c$ (the normal state). For MMP and ZBP, $T_x$ is the temperature below which $\xi(T)$ becomes constant. Multiplying the MMP $\mathrm{Im}\chi(\mathbf{q},\omega)/\omega$ by the $T$ factor in the first term on the right-hand side of equation~\ref{moriyaexp} leads to our expression in equation ~\ref{tauaf} for $\tau_{AF}$.

In their fit to the NMR, MMP found $T_x\sim 100$ K~\cite{MMP} (ZBP found $T_x=157$ K~\cite{ZBP,Nandor1999}). MMP commented that, {\it ``We found in Section III that the increase of $\xi$ as $T$ decreases is cut off on the surprisingly low energy scale of $T_x\sim 100$ K. ... However, it is puzzling that such a small energy scale should appear in a material where the basic electronic energy scales ... are of the order of several tenths of a volt and where the correlations lengths are not too long. ... It is not clear to us why, in the absence of a magnetic transition, the correlation length should so abruptly saturate. Further, it is remarkable that the antiferromagnetic corrrelations whose characteristic energy scale is\ $T_x\sim 100$ K are apparently unaffected by the onset of superconductivity at $T_c\sim 90$ K. This issue is discussed further by one of us."} A citation to an unpublished paper by Pines is placed at the end of this quote. I have not located this discussion by Pines.

In our model, $T_x$ is the energy scale for the inter-AF-cluster spin-spin coupling. It is expected to be small. In fact, we will show later that $T_x\approx 60$ K. Also, the AF spin correlation length can never become larger than the average size of the discrete AF clusters (a few lattice spacings). Thus, the temperature where the correlation length inside each cluster maximizes is well above room-temperature. Below this temperature, the spin correlation length seen by neutron scattering, for example, will remain constant. The discrete AF clusters in our model provide explanations for both the small $T_x$ and the small saturated spin correlation length.

The $T$ dependence of the static magnetic spin susceptibility from the AF clusters up to $700$ K (less than $1/2$ of $J_{AF}\approx 1500$ K) should be Curie-like with an energy scale set by $J_{AF}$. The static spin susceptibility of an isolated spin (Curie paramagnetism) is

\begin{equation}
\label{chicurie}
\chi_{\mathrm{Curie}}(T)=\frac{(g\mu_B)^2}{4}\left(\frac{1}{k_B T}\right).
\end{equation}

The AF static susceptibility should scale the same way as equation~\ref{chicurie} for $T>>T_x$. Therefore, we choose the form for the $T$ dependence of the AF static magnetic susceptibility to be

\begin{equation}
\label{chiAF}
\chi_{AF}(T) = \kappa\mu_B^2\left[\frac{1}{k_B(T + T_x)}\right],
\end{equation}

\noindent where $\kappa$ is a dimensionless constant. The functional form of the temperature dependence of $\chi_{AF}(T)$ is the same as $\tau_{AF}/T$ from equation~\ref{tauaf}. In equation~\ref{chicurie}, $\kappa=(1/4)g^2=1$. For an AF cluster, the total magnetization is the sum of all the spins in the cluster. Since the AF spin correlation length, $\xi$, in the cluster is approximately the size of the cluster, $\kappa$ should be much smaller than its value for the single unpaired spin in the Curie susceptibility. Our fits to the NMR below find $\kappa\approx 0.05$.

\subsection{Metallic Region Spin Correlation Time and Static Magnetic Susceptibility}

Inside the metallic region, the electron spin correlation time is

\begin{equation}
\label{taum0}
\tau_{M}\sim\left(\frac{k_B T}{E_F}\right)\left(\frac{\hbar}{E_F}\right),
\end{equation}

\noindent where $E_F$ is the Fermi energy. The $\hbar/E_F$ term is the dwell time and the $(k_B T/E_F)$ term is the fraction of the time that dephasing occurs~\cite{PinesSlichter}.

The simple band theory expression is

\begin{equation}
\label{taum1}
\tau_{\mathrm{band},M}(T)=2\pi\hbar N(0)^2(k_B T),
\end{equation}

\noindent where $N(0)$ is the metallic density of states in units of states per spin per energy per planar Cu. Since $N(0)\sim 1/E_F$, the only difference between equations~\ref{taum0} and ~\ref{taum1} are the constants in ~\ref{taum1}.

Fluctuations of the metallic Cu $d_{x^2-y^2}$ orbital energy occur due to coupling to nearby AF localized Cu spins. These AF fluctuations may raise the Cu $d$ orbital energy and simultaneously lower it at a neighboring metallic Cu $d$ orbital, or vice versa. Hence, it takes longer for an electron to ``hop" to a neighboring site compared to the hoppping time when the neighboring orbital energies are equal. Thus, the right-hand side of equation~\ref{taum1} should be multiplied by a dimensionless ``dwell time enhancement factor," $\lambda_{dwell}$. Equation~\ref{taum1} then becomes

\begin{equation}
\label{taum}
\tau_{M}(T)=2\pi\hbar N(0)^2(k_B T)\cdot\lambda_{dwell},
\end{equation}

The dimensionless $\lambda_{dwell}$ comes from fluctuations of the metallic Cu $d_{x^2-y^2}$ orbital energy due to coupling to the AF clusters localized Cu spins. No dwell time enhancement is $\lambda_{dwell}=1$. We find $\lambda_{dwell}\approx 2.5$ for cuprates.

In equation~\ref{taum}, we assumed that the matrix element for scattering an electron from a state at the Fermi level to another state at the Fermi level is constant. Including Fermi surface effects into this equation leads to averages of $\cos k_x a$, $\cos^2 k_x a$, $\cos k_y a$, $\cos^2 k_y a$, and $\cos k_x a\cos k_y a$ over the Fermi surface where $a$ is the $\mathrm{CuO_2}$ lattice size, $k_x$ is the electron momentum along the x-axis (along one of the planar Cu-O bond directions), and $k_y$ is the planar momentum perpendicular to the x-axis. These corrections are of order $\sim 1$.

For the metallic region, the static magnetic spin susceptibility, $\chi_M$, is independent of temperature and is given by

\begin{equation}
\label{chiM}
\chi_M=\frac{1}{2}(g\mu_B)^2N(0)=\mu_B^2 N_{\mathrm{tot}}(0),
\end{equation}

\noindent where  $N_\mathrm{tot}=2N(0)$ is the total density of states.

Equations~\ref{tauaf}, ~\ref{chiAF}, ~\ref{taum}, and ~\ref{chiM} define our expressions for the metallic and AF regions spin correlation times and static susceptibilities. The parameters that may be adjusted to fit the NMR data are $T_x$, $\tau_0$, $N(0)$, and $\lambda_{dwell}$.

\subsection{Temperature Independent AF Spin Correlation Length in the AF Region}

The AF spin correlation length in each discrete AF cluster is approximately the cluster size for temperatures below our maximum temperature of $700$ K. Thus we assume there is a temperature independent AF spin correlation length, denoted by $\xi_{AF}$, that leads to a static spin correlation in the AF clusters. Following Uldry and Meier~\cite{Uldry2005}, we define the normalized ``static" spin correlation as

\begin{equation}
\label{staticcorr}
K^\alpha(\mathbf{R})=\frac{<S^\alpha(\mathbf{R})S^\alpha(\mathbf{0})>}{<(S^\alpha)^2>}=\Theta(\mathbf{R}) e^{-R/\xi_{AF}}
\end{equation}

\noindent The vector $\mathbf{R}$ is the location vector for the first spin relative to the second spin at location vector $\mathbf{0}$. Since the electron spin is $1/2$, the magnitude of the square of the electron spin component is $<(S^\alpha)^2> = (1/2)^2=1/4$. Due to the AF spin-spin coupling, $\Theta(\mathbf{R})=+1$, if $\mathbf{R}$ is an even number of lattice steps from $\mathbf{0}$, and $\Theta(\mathbf{R})=-1$, if $\mathbf{R}$ is an odd number of lattice steps from $\mathbf{0}$.

Therefore, the exponential time decay of an AF spin correlation is~\cite{Uldry2005}

\begin{equation}
\label{timecorr}
\frac{\left<S^\alpha(\mathbf{R},t)S^\alpha(\mathbf{0})\right>}{\left<(S^\alpha)^2\right>}=K^\alpha(\mathbf{R})e^{-t/\tau_{AF}},
\end{equation}

\noindent where $\tau_{AF}$ is defined in equation~\ref{tauaf}.

The electronic AF spin correlations should be independent of the spin component direction $\alpha$. Hence, $K^\alpha$ is independent of $\alpha$. This $\alpha$ independence was not found by Uldry and Meier~\cite{Uldry2005} in their model for the spin relaxation up to $300$ K. They suggested that the spin component direction dependence may be due to anisotropies in g-factors. However, these effects are likely small for the antiferromagnetism in our AF clusters. Therefore, we believe $K^\alpha$ should be independent of $\alpha$. Henceforth, we drop the superscript $\alpha$.

\begin{figure}[tbp]
\centering \includegraphics[width=0.5\linewidth]
{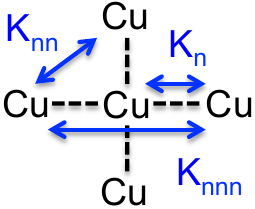}
\caption{Definitions of the static AF spin correlations, $K_{n}$ (neighbor), $K_{nn}$ (next-neighbor), and $K_{nnn}$ (next-next-neighbor) as defined by equation~\ref{staticcorr}. Due to AF spin correlation, $K_{n}<0$, $K_{nn}>0$, and $K_{nnn}>0$.
}
\label{nnn}
\end{figure}

There are three $\mathbf{R}$ separations that are relevant to the NMR. They are to the neighboring Cu, $K_n$, the next-nearest neighbor, $K_{nn}$, and the next-next-nearest Cu neighbor, $K_{nnn}$ as shown in Figure~\ref{nnn}.

\subsection{Electron-Nuclear Hyperfine Couplings}

Just like MMP and ZBP, we assume there are electron spins at the planar Cu sites with components $S^x$, $S^y$, and $S^z$ where the $x$ and $y$ axes are along the Cu-O bond directions in the $\mathrm{CuO_2}$ planes (also called the $a$ and $b$-axes, respectively) and $z$ is normal to the $\mathrm{CuO_2}$ planes (also called the $c$-axis). We label a planar Cu atom (or nucleus) by the unit cell index, $n$, where it resides.

The Cu atom at $n$ is either a metallic Cu (in the metal region) or an AF Cu (in the antiferromagnetic region). The electronic states for a metallic Cu are delocalized and the electronic states for an AF Cu are localized spins. The electron-nuclear Hamiltonian for Cu, O, and Y nuclei in unit cell $n$ is the sum of the couplings to the neighboring metallic Cu atoms and AF Cu atoms,

\begin{equation}
\label{hamsum}
H_{e-n}=H_{e-n}^{AF} + H_{e-n}^M,
\end{equation}

\noindent where the total, $H_{e-n}$, is the sum of the electron-nuclear couplings $H_{e-n}^{AF}$ and $H_{e-n}^M$, where the former is the coupling to the AF Cu electronic spins and the latter is the coupling to the metallic Cu electronic spins. $H_{e-n}$ is expanded mathematically in equations~\ref{hamaf} and ~\ref{hamm} below. They are shown schematically in figures~\ref{Cucoup} and ~\ref{OYcoup}.

\begin{figure}[tbp]
\centering \includegraphics[width=0.9\linewidth]
{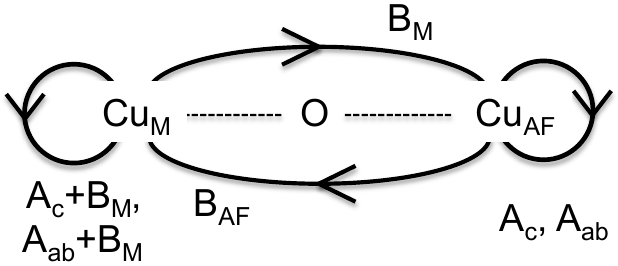}
\caption{Onsite and transferred hyperfine couplings of two neighboring planar Cu nuclei, one metallic and one AF, as described by equations~\ref{hamsum}, ~\ref{hamaf}, and ~\ref{hamm}. The subscript on the Cu atom denotes the region the Cu atom comes from (Metal or AF). The O atom in-between the two Cu atoms is shown. There are onsite hyperfine couplings, $A_c$ and $A_{ab}$, of the $d_{x^2-y^2}$ orbital spins at each Cu atom to its nucleus. At the metallic $\mathrm{Cu_M}$, the $d_{x^2-y^2}$ and $4s$ orbital couplings sum. In the figure, it is represented by the sums $A_c+B_M$ and $A_{ab}+B_M$. These expressions are merely symbolic shortcuts for the expressions in equations~\ref{phiab} and ~\ref{phic}. The Cu $d_{x^2-y^2}$ character at $\mathrm{Cu_M}$ leads to Cu $4s$ character at the AF Cu, or $\mathrm{Cu_{AF}}$. This $4s$ character at $\mathrm{Cu_{AF}}$ leads to the transferred hyperfine coupling $B_M$ shown above the arrow in the top of the figure. The bottom arrow in the figure shows the transferred hyperfine coupling $B_{AF}$ from the AF Cu atom on the right to the metallic Cu atom on the left and its $4s$ orbital that arises from the Cu $d_{x^2-y^2}$ orbital at $\mathrm{Cu_{AF}}$.
}
\label{Cucoup}
\end{figure}

\begin{figure}[tbp]
\centering \includegraphics[width=0.7\linewidth]
{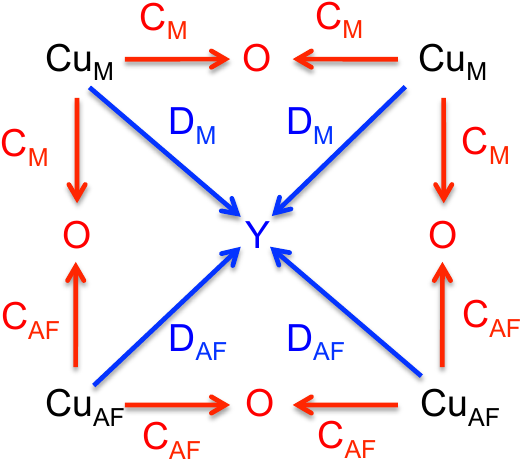}
\caption{Oxygen and Yttrium hyperfine couplings of four planar Cu nuclei, two metallic and two AF, as described by equations~\ref{hamsum}, ~\ref{hamaf}, and ~\ref{hamm}. The subscript on the Cu atom denotes the region the Cu atom comes from (Metal or AF).
}
\label{OYcoup}
\end{figure}

The coupling to AF Cu spins is identical to the MMP~\cite{MMP} Hamiltonian

\begin{eqnarray}
\label{hamaf}
H_{e-n}^{AF} & = & \delta_{n,AF}\left[A_{ab,AF}(^{63}I_n^x S_n^x +{^{63}I_n^y}S_n^y) + A_{c,AF}{^{63}I_n^z}S_n^z\right] \nonumber \\
& + & B_{AF}\sum_{n'=AF}\mathbf{^{63}I_n\cdot S_{n'}} + C_{AF}\sum_{n'=AF}\mathbf{^{17}I_n\cdot S_{n'}} \nonumber \\
& + & D_{AF}\sum_{n'=AF}\mathbf{^{89}I_n\cdot S_{n'}}.
\end{eqnarray}

\noindent Here, $^{k}I_n^\alpha$ is the $\alpha$ component of the nuclear spin isotope with atomic number $k$ in unit cell $n$. $S_n^\alpha$ is the $\alpha$ component of the electron spin at the Cu site in unit cell $n$. The delta function, $\delta_{n,AF}$, equals one when the Cu site in unit cell $n$ is an AF Cu site and equals zero if the Cu atom at $n$ is metallic. Similarly, $n'=AF$ in the summation means $\delta_{n',AF}$. The sums of $n'$ are over the nearest neighbor Cu sites to unit cell $n$. $A_{ab}$ is the onsite planar hyperfine coupling and $A_c$ is the onsite coupling normal to the plane arising from the Cu $d_{x^2-y^2}$ orbital on the Cu site. $B_{AF}$ is the isotropic transferred hyperfine coupling. $C_{AF}$ and $D_{AF}$ are the isotropic couplings to the O and Y atoms, respectively. All the hyperfine coupling constants have units of energy.

The coupling to metallic Cu spins is

\begin{eqnarray}
\label{hamm}
H_{e-n}^{M} & = & \delta_{n,M}\left[\mathbf{\Phi_{ab,n}}(^{63}I_n^x S_n^x +{^{63}I_n^y}S_n^y) + \mathbf{\Phi_{c,n}}{^{63}I_n^z}S_n^z\right] \nonumber \\
& + & B_{M}\sum_{\substack{n=AF \\ n'=M}}\mathbf{^{63}I_n\cdot S_{n'}} + C_{M}\sum_{n'=M}\mathbf{^{17}I_n\cdot S_{n'}} \nonumber \\
& + & D_{M}\sum_{n'=M}\mathbf{^{89}I_n\cdot S_{n'}}.
\end{eqnarray}

\noindent where $\mathbf{\Phi_{ab,n}}$ and $\mathbf{\Phi_{c,n}}$ are the metallic band orbital operators

\begin{equation}
\label{phiab}
\mathbf{\Phi_{ab,n}}=A_{ab,M}'|\varphi_{d,n}><\varphi_{d,n}| + B_{M}'|\varphi_{4s,n}><\varphi_{4s,n}|,
\end{equation}

\begin{equation}
\label{phic}
\mathbf{\Phi_{c,n}}=A_{c,M}'|\varphi_{d,n}><\varphi_{d,n}| + B_{M}'|\varphi_{4s,n}><\varphi_{4s,n}|.
\end{equation}

\noindent The wavefunctions $\varphi_{d,n}$ and $\varphi_{4s,n}$ are the Cu $d_{x^2-y^2}$ and Cu $4s$ orbitals at site $n$, respectively. $C_M$ and $D_M$ are the isotropic O and Y atom couplings, respectively. The parameters $A_{c,M}'$ and $A_{ab,M}'$ are the hyperfine couplings of a $d_{x^2-y^2}$ orbital to its nucleus. $B_{M}'$ is the hyperfine coupling of a $4s$ orbital to its nucleus. All these parameters have units of energy.

For spin relaxation, matrix elements between different delocalized band states near the Fermi level in equation~\ref{hamm} are squared and then averaged over the Fermi surface. Since the $d_{x^2-y^2}$ and $4s$ orbitals have different symmetry under $90^{\circ}$ rotations, the cross term $\sim A'B'$ cancels out in the average over the Fermi surface~\cite{Obata1963,Yafet1964}. The $A'^{2}$ and $B'^{2}$ terms are multiplied by the square of the fraction of $d_{x^2-y^2}$ and $4s$ orbital character at site $n$ over the Fermi surface, or $<|\varphi_{d,n}|^2>_{FS}$ and $<|\varphi_{4s,n}|^2>_{FS}$. The total orbital character at $n$ over the Fermi surface sums to one,

\begin{equation}
\label{orbsum}
\left<|\varphi_{d,n}|^2\right>_{FS}+\left<|\varphi_{4s,n}|^2\right>_{FS}=1.
\end{equation}

\noindent Hence, what we call $A_{ab,M}$, $A_{c,M}$, and $B_M$ corresponding to the $A_{ab,AF}$, $A_{c,AF}$, and $B_{AF}$ in equation~\ref{hamaf} are the products

\begin{eqnarray}
\label{coupfs}
A_{ab,M} & = & A_{ab}'\left<|\varphi_{d,n}|^2\right>_{FS},\nonumber \\
A_{c,M}  & = & A_{c}'\left<|\varphi_{d,n}|^2\right>_{FS},\nonumber \\
B_{M}    & = & B_{M}'\left<|\varphi_{d,n}|^2\right>_{FS}.
\end{eqnarray}

In the superconducting state near $T=0$, the superconducting d-wave gap will limit electron spin-flip scatterings across the Fermi surface to states near the nodes of the gap (diagonals in the Brillouin Zone). Along the Brillouin Zone diagonals, the metallic states have no Cu $4s$ character because $d_{x^2-y^2}$ and $4s$ transform differently under diagonal $\mathbf{k}$ vector reflections. The diagonal states are purely Cu $d_{x^2-y^2}$. Hence, the $B_M'$ terms in equations~\ref{phiab} and ~\ref{phic} vanish leading to a large increase in the Cu spin relaxation rate anisotropy at very low temperatures. This issue is discussed later when we analyze the results of our model in the superconducting state.

Since the metallic band is half-filled (one metallic electron per Cu) and the AF region has one electron per AF Cu site, the metallic and AF Cu $d_{x^2-y^2}$ hyperfine couplings should be approximately equal. In this paper, we assume they are equal leading to

\begin{equation}
\label{avals}
A_{c,AF}=A_{c,M}\equiv A_c,\ \ \ A_{ab,AF}=A_{ab,M}\equiv A_{ab}.
\end{equation}

\noindent In general, $B_{M}\ne B_{AF}$. The fits in this paper are for $B_{M}=B_{AF}$.

Since the nuclear magnetic moment for the isotope $\mathrm{^{63}Cu}$ is different from its isotope $\mathrm{^{65}Cu}$, the values of hyperfine couplings, $A_{c}$, $A_{ab}$, and $B$, are different for each isotope. The ratio of their hyperfine couplings, $^{65}A_c/^{63}A_c$ for example, is equal to the ratio of their nuclear magnetic moments, $^{65}\gamma/^{63}\gamma\approx 1.07$. Our fitted values for these hyperfine couplings are quoted for $\mathrm{^{63}Cu}$ only.

In ZBP, the Cu spin to O isotropic nuclear hyperfine constant, $C$, was separated into $C_a$ and $C_b$ for planar spin components and $C_c$ for the spin component normal to the planes in order to incorporate the small difference in the spin relaxation rate for the three magnetic field directions. Since we are fitting to the c-axis O relaxation data of ~\cite{Nandor1999}, we assume $C_a=C_b=C_c=C$ for simplicity. This assumption was also made by MMP. Finally, ZBP had a next-nearest-neighbor hyperfine coupling to the O nucleus, $C'$. This parameter was invoked to fit the NMR data with incommensurate spin fluctuations, as seen by neutron scattering. Here, we set $C'=0$.

\subsection{Normal State Spin Relaxation Rate Expressions}

Atomic-scale inhomogeneity with a metallic region and an AF region leads to several different environments, or configurations, for the nuclei. There is a different spin relaxation rate and Knight shift for each configuration. We assume that there is rapid energy exchange between identical nuclei such that they obtain the same spin temperature~\cite{Slichterbook}. With a spin temperature, the observed spin relaxation rate and Knight shift are an average of the individual nuclear relaxations and shifts.

The spin relaxation rate, $^{k}(1/T_1)_\alpha$, for nuclear isotope $k$ and magnetic field along the $\alpha$ direction is due to fluctuating magnetic fields in the two directions perpendicular to the direction $\alpha$. The relaxation rate from a fluctuating magnetic field in the $\beta$ direction is denoted by $^{k}U_\beta$ (we adopt the notation of Uldry and Meier~\cite{Uldry2005}) leading to

\begin{equation}
\label{t1fluc}
^{k}(1/T_1)_\alpha = \sum_{\beta\ne\alpha}{^{k}U_\beta}.
\end{equation}

With both metallic electrons and AF electrons, $^{k}U_\beta$ is the sum of two terms,

\begin{equation}
\label{utot}
^{k}U_\beta=^{k}U_{\beta,M}+^{k}U_{\beta,AF},
\end{equation}

\noindent where $^{k}U_{\beta,M}$ are $^{k}U_{\beta,AF}$ are the relaxation rates from the  metallic and AF regions, respectively.

Each $^{k}U_\beta$ term is the Fourier transform at the nuclear Larmor frequency of the time correlation of its fluctuating magnetic field in the $\beta$ direction. The result is~\cite{Slichterbook},

\begin{equation}
\label{Ueqn}
^{k}U_{\beta,R} = \left(\frac{1}{\hbar^2}\right)\left<h_{\beta,R}^2\right>\tau_{c,R},\ \ \ R\in\{M,AF\},
\end{equation}

\noindent where the subscript $R$ labels the region ($R=M$ or $R=AF$), $\tau_{c,R}$ is the region's electron spin correlation time, and $<h_{\beta,R}^2>$ is the mean of the square of the fluctuating magnetic field magnitude of the region. In this form, $h_{\beta,R}$ has units of energy.

In general,

\begin{equation}
\label{fluch}
h_{\beta,R}=\sum_{n\in R} A_{\beta,n}S^{\beta}_n,
\end{equation}

\noindent is sum of the $\beta$ component of the spins at the sites $n$ in region $R$ that couple to the nucleus multiplied by their hyperfine couplings, $A_\beta$. Squaring this equation leads to

\begin{equation}
\label{hsqm}
\left<h_{\beta,R}^2\right>=\sum_{n_1,n_2\in R} A_{\beta,n_1}A_{\beta,n_2}\left<S^\beta_{n_1}S^\beta_{n_2}\right>.
\end{equation}

The effective square hyperfine interaction, $A_{\mathrm{eff},R}^2$, is

\begin{equation}
\label{Aeff}
A_{\mathrm{eff},R}^2\equiv \frac{\left<h_{\beta,R}^2\right>}{\left<(S^\beta)^2\right>}= 4\sum_{n_1,n_2\in R} A_{\beta,n_1}A_{\beta,n_2}\left<S^\beta_{n_1}S^\beta_{n_2}\right>.
\end{equation}

Let $C$ be a configuration surrounding a nucleus and $P(C)$ its probability. Multiplying the probability of a configuration by its $^{k}U_\beta(C)$ and summing over all configurations leads to

\begin{eqnarray}
\label{usumm}
^{k}U_{\beta,M} & = & \left(\frac{1}{2}\right)^2\left(\frac{1}{\hbar^2}\right)\tau_M \sum_{C}P(C)A_{\mathrm{eff},M}^2(C), \\
\label{usumaf}
^{k}U_{\beta,AF} & = & \left(\frac{1}{2}\right)^2\left(\frac{1}{\hbar^2}\right)\tau_{AF} \sum_{C}P(C)A_{\mathrm{eff},AF}^2(C).
\end{eqnarray}

\subsubsection{Cu Spin Relaxation}

\begin{figure}[tbp]
\centering \includegraphics[width=0.9\linewidth]
{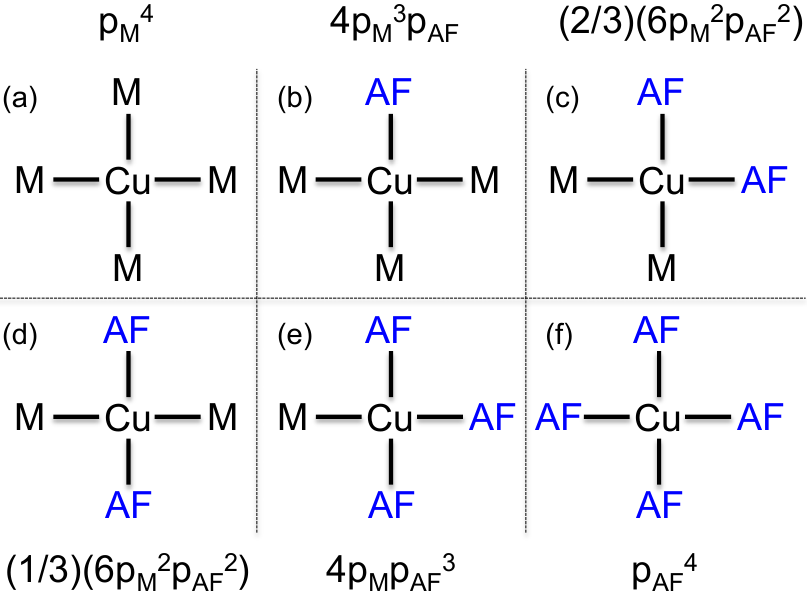}
\caption{The six topologically distinct configurations of metal and AF Cu atoms neighboring a Cu atom (metal or AF). A metal neighbor is denoted by a capital M and an AF neighbor is denoted by AF in blue. The probability for each configuration is shown above or below each smaller figure. $p_M$ is the probability that a Cu atom is in the metallic region, and $p_{AF}$ is the probability the atom is in the AF region. (a) Four metal neighbors with probability $p_M^4$. (b) Three metal neighbors and one AF neighbor with probability $4p_M^3p_{AF}$. (c) and (d) Two metal and two AF neigbors with probability $(2/3)(6p_M^2p_{AF}^2)=4p_M^2p_{AF}^2$ when the AF sites are closest in (c), and probability $(1/3)(6p_M^2p_{AF}^2)=2p_M^2p_{AF}^2$ in (d) when the two AF sites are across the central Cu atom. (e) One metal and three AF neighbors with probability $4p_M p_{AF}^3$. (f) Four AF neighbors with probability $p_{AF}^4$.
}
\label{Cuneigh}
\end{figure}

A Cu nucleus may be either inside a metallic Cu atom with probability $p_M$ or inside an AF Cu atom with probability $p_{AF}$. In addition, the nucleus may be surrounded by $n_M$ metallic Cu neighbors and $n_{AF}$ AF Cu neighbors where $n_M+n_{AF}=4$. $n_M$ varies from $0$ to $4$.

Figure~\ref{Cuneigh} shows the six topologically distinct metal and AF Cu neighbors surrounding a Cu atom with their respective probabilities. There are two distinct configurations when $n_M=n_{AF}=2$ as shown in Figures~\ref{Cuneigh}(c) and \ref{Cuneigh}(d). Since the Cu atom containing the nucleus can be metallic or AF, there are a total of $2\times 6=12$ configurations in the sum of equation~\ref{usumm} for $^{k}U_{\beta,M}$, and $12$ configurations in the sum of equation~\ref{usumaf} for $^{k}U_{\beta,AF}$.

The probability for one configuration is the product of the probability of the type of the Cu atom containing the nucleus ($p_M$ for metal and $p_{AF}$ for AF) times the probability of the particular Cu neighbor configuration shown in Figure~\ref{Cuneigh}. For example, the probability of an AF nucleus with 3 metal and 1 AF neighbors is $p_{AF}(4p_M^3 p_{AF})$ [see Figure~\ref{Cuneigh}(b)].

Using equation~\ref{Aeff}, $A_{\mathrm{eff},AF}^2$ equals,

\begin{eqnarray}
\label{Aaf0}
A_{\mathrm{eff},AF}^2[\mathrm{Fig~\ref{Cuneigh}(a)}] & = & A_\beta^2, \\
A_{\mathrm{eff},AF}^2[\mathrm{Fig~\ref{Cuneigh}(b)}] & = & A_\beta^2+1\left( B_{AF}^2 + 2K_{n}A_\beta B_{AF}\right), \\
A_{\mathrm{eff},AF}^2[\mathrm{Fig~\ref{Cuneigh}(c)}] & = & A_\beta^2+2\left( B_{AF}^2 + 2K_{n}A_\beta B_{AF}\right) \nonumber \\
&  & + 2K_{nn}B_{AF}^2, \\
A_{\mathrm{eff},AF}^2[\mathrm{Fig~\ref{Cuneigh}(d)}] & = & A_\beta^2+2\left( B_{AF}^2 + 2K_{n}A_\beta B_{AF}\right) \nonumber \\
&  & + 2K_{nnn}B_{AF}^2, \\
A_{\mathrm{eff},AF}^2[\mathrm{Fig~\ref{Cuneigh}(e)}] & = & A_\beta^2+3\left( B_{AF}^2 + 2K_{n}A_\beta B_{AF}\right) \nonumber \\
&  & + 2(2K_{nn}+K_{nnn})B_{AF}^2, \\
\label{Aaf4}
A_{\mathrm{eff},AF}^2[\mathrm{Fig~\ref{Cuneigh}(f)}] & = & A_\beta^2+4\left( B_{AF}^2 + 2K_{n}A_\beta B_{AF}\right) \nonumber \\
&  & + 2(4K_{nn}+2K_{nnn})B_{AF}^2.
\end{eqnarray}

For configurations where the Cu atom containing the nucleus is AF, $A_\beta=A_c$ or $A_{ab}$ depending on the value of $\beta$. If the nucleus resides inside a metallic Cu atom, $A_\beta=0$.

From equation~\ref{Aeff}, $A_{\mathrm{eff},M}^2$ is

\begin{equation}
\label{Am}
A_{\mathrm{eff},M}^2=A_\beta^2 + n_M\cdot B_M^2,
\end{equation}

\noindent where $n_M$ is the number of metal neighbors in figure~\ref{Cuneigh}. $A_\beta=A_c$ or $A_{ab}$ if the nucleus is in a metallic Cu atom, and $A_\beta=0$ if the nucleus is in an AF Cu atom. The cross term, $A_\beta B_M$, does not appear because its coeffecient averages to zero over the Fermi surface since $d_{x^2-y^2}$ and $4s$ orbitals have different symmetries under $90^{\circ}$ rotation.

Substituting the probabilities in Figure~\ref{Cuneigh}, the equations~\ref{Aaf0}$-$\ref{Am} for the effective hyperfine couplings squared, and the two equations for the spin correlation times in the AF and metal regions, \ref{tauaf} and \ref{taum}, into equations~\ref{usumm} and \ref{usumaf} leads to $^{63}(1/T_1)$.

\subsubsection{O Spin Relaxation}

\begin{figure}[tbp]
\centering \includegraphics[width=0.9\linewidth]
{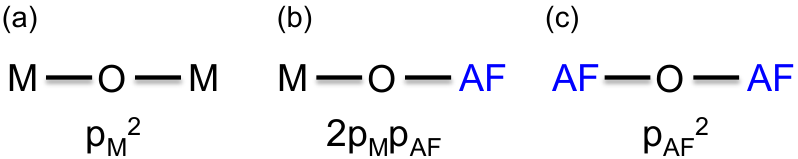}
\caption{The three topologically distinct configurations of metal and AF Cu atoms neighboring a O atom. A metal neighbor is denoted by a capital M and an AF neighbor is denoted by AF in blue. The probability for each configuration is shown below each smaller figure. (a) Two metal neighbors with probability $p_M^2$. (b) One metal and one AF neighbor with probability $2p_M p_{AF}$. (c) Two AF neigbors with probability $p_{AF}^2$.
}
\label{Oneigh}
\end{figure}

Figure~\ref{Oneigh} shows the three topologically distinct metal and AF Cu neighbors surrounding an O atom. Their respective probabilities are shown below each configuration.

The effective hyperfine couplings squared for the AF region electrons is,

\begin{eqnarray}
\label{OxyAaf0}
A_{\mathrm{eff},AF}^2[\mathrm{Fig~\ref{Oneigh}(a)}] & = & 0, \\
\label{OxyAaf1}
A_{\mathrm{eff},AF}^2[\mathrm{Fig~\ref{Oneigh}(b)}] & = & C_{AF}^2, \\
\label{OxyAaf2}
A_{\mathrm{eff},AF}^2[\mathrm{Fig~\ref{Oneigh}(c)}] & = & 2C_{AF}^2(1+K_{n}).
\end{eqnarray}

\noindent The effective hyperfine coupling for the metal region for $n_M$ metal neighbors to the O atom is

\begin{equation}
\label{OxyAm}
A_{\mathrm{eff},M}^2 = n_{M}\cdot C_M^2.
\end{equation}

Substituting the probabilities in Figure~\ref{Oneigh}, the equations~\ref{OxyAaf0}$-$\ref{OxyAm} for the effective hyperfine couplings squared, and the two equations for the spin correlation times in the AF and metal regions, \ref{tauaf} and \ref{taum}, into equations~\ref{usumm} and \ref{usumaf} leads to $^{17}(1/T_1)$.

\subsubsection{Y Spin Relaxation}

The observation by Nandor et al~\cite{Nandor1999} quoted in the introduction that the Yttrium spin relaxation rate divided by temperature, $^{89}(1/T_1 T)$, is constant from $300$ to $700$ K (within $\pm 4\%$) while $\chi(T)$ drops $\approx 13\%$ over the same temperature range cannot be explained by the model we have developed so far because, just like MMP and ZBP, our $^{89}(1/T_1 T)$ must ``track" $\chi(T)$. Therefore, there is a missing piece of physics that is necessary to explain Nandor's data.

We believe the missing piece is that the transverse $T_2$ relaxation time of the Y nucleus due to metallic electrons is larger than the $T_2$ relaxation time of the AF electrons.

Any experiment that creates an ``echo train" by repeated $\pi$ pulses will find that the echo signal amplitude from the AF and metal regions, as a function of time, $t$, decay as,

\begin{equation}
I_{AF}(t)\sim e^{-t/T_{2,AF}},\ \ \ I_{M}(t)\sim e^{-t/T_{2,M}}.
\end{equation}

If $T_{2,AF}<<T_{2,M}$, then the AF spin relaxation component will decay away faster then the M spin relaxation component. Over a long echo train, the spin relaxation rate will converge to the metallic relaxation. Since a Knight shift measurement performs one $\pi/2$ pulse followed by a $\pi$ pulse before recording the echo and the time interval between the two pulses is short, the Knight shift will record the total response from both regions.

The linewidth of the Y nucleus at $293$ K is $\sim 3-4$ times larger~\cite{Alloul1988} in the parent AF material, $\mathrm{YBa_2Cu_3O_6}$, compared to fully oxygenated superconducting $\mathrm{YBa_2Cu_3O_7}$. This observation suggests the $T_2$ in the AF region will be shorter than $T_2$ in the metallic region and lends support to our assumption that $T_{2,AF}<<T_{2,M}$.

In order to increase the signal-to-noise during their $^{89}(1/T_1)$ measurement, Nandor et al used a Carr-Purcell-Meiboom-Gill (CPMG)~\cite{Slichterbook,CP,MG} pulse sequence to refocus the echo after a saturation pulse. They created an echo train with 256 echoes using $\pi$ pulses. They found that the 50th echo in the echo train had an amplitude that was approximately $15\%$ of the amplitude of the initial echo. The full echo train amplitude was weighted by an exponential in order to extract $^{89}(1/T_1)$.

We believe the CPMG pulse sequence suppressed the AF region contribution to the Y relaxation and led to the metallic electron contribution dominating over the AF region contribution. The Knight shift data was not collected on a CPMG echo train. Hence, the shift captured its full AF contribution. Since the metallic region leads to a linear $T$ contribution to $^{89}(1/T_1)$, the discrepancy between the temperature dependencies from $300-700$ K of $^{89}(1/T_1 T)$ and $^{89}K$ is explained.

The effective AF hyperfine coupling to Y squared is therefore zero and the metallic effective coupling is,

\begin{equation}
\label{YAeff}
A_{\mathrm{eff},AF}^2=0,\ \ \ A_{\mathrm{eff},M}^2=n_M\cdot D_M^2,
\end{equation}

\noindent where $n_M$ is the number of metal neighbors to the Y atom. Equation~\ref{YAeff} allow us to calculate $^{89}(1/T_1)$.

\subsection{Static Susceptibility, O Knight Shift, and Y Knight Shift}

The static magnetic susceptibility is given by the average of the AF and metallic region susceptibilities (equations~\ref{chiAF} and \ref{chiM}),

\begin{equation}
\label{chif}
\chi(T)=p_M\cdot\chi_{M}(T) + p_{AF}\cdot\chi_{AF}(T).
\end{equation}

The O Knight shift is,

\begin{equation}
\label{KO}
^{17}K_c= \frac{2\left[p_M\cdot C_M\chi_{M} + p_{AF}\cdot C_{AF}\chi_{AF}\right]}{{(^{17}\gamma\hbar)(\gamma_e\hbar)}},
\end{equation}

\noindent where the factor of $2$ in the numerator is the number of Cu neighbors of an O atom, $^{17}\gamma$ is the gyromagnetic ratio of $\mathrm{^{17}O}$, and $\gamma_e\hbar=g\mu_B$ where $\gamma_e$ is the gyromagnetic ratio of the electron.

The Y Knight shift is,

\begin{equation}
\label{KY}
^{89}K_c= \frac{8\left[p_M\cdot D_M\chi_{M} + p_{AF}\cdot D_{AF}\chi_{AF}\right]}{{(^{89}\gamma\hbar)(\gamma_e\hbar)}},
\end{equation}

\noindent where the factor of $8$ in the numerator is the number of Cu neighbors of an Y atom (4 per $\mathrm{CuO_2}$ plane) and $^{89}\gamma$ is the gyromagnetic ratio of $\mathrm{^{89}Y}$.

\section{Choice of NMR Data to Fit for \texorpdfstring{$\mathbf{YBa_2Cu_3O_7}$}{YBCO}}

Normal state spin relaxation data for the planar Cu, the planar O, and the Y atoms in YBa$_2$Cu$_3$O$_7$ and the static magnetic susceptibility are used to fit the model in this paper. The fitted model is then tested against the temperature dependences of the O and Y Knight shifts. We wanted data that went up the to highest possible temperatures and had measured the spin relaxation for Cu with the magnetic field in the CuO$\mathrm{_2}$ (ab-axis magnetic field) and perpendicular to the CuO$\mathrm{_2}$ plane (c-axis magnetic field).

Walstedt et al~\cite{Walstedt1989} measured both the c-axis and ab-axis relaxation rates as shown in Figure 2 of their paper up to $300$ K.
Barrett et al~\cite{Barrett1991} measured the c-axis $^{63}(1/T_1)$ up to $500$ K. This data is shown in Figure 4 of their paper and comes from NQR on an unaligned powder sample (they call it Sample 3). We multiply Barrett's relaxation rate, $W_{1c}$ by a factor of $2/3$ to obtain $^{63}(1/T_1)_c$.

For the c-axis Cu data, we use Barrett et al~\cite{Barrett1991} that goes up to $500$ K. We use the ab-axis data of Walstedt et al~\cite{Walstedt1989} that goes up to $300$ K. The reason we do not use the ab-axis data of Barrett et al~\cite{Barrett1991} is because it is composed of a total of 5 data points with two very close together at $\approx$ 100 K, two very close together at $\approx$ 130 K, and one data point at $\approx$ 300 K. Hence, it is really 3 data points. On the other hand, Walstedt et al~\cite{Walstedt1989} has 9 data points spread more uniformly from $\mathrm{100-300}$ K.

The c-axis Oxygen spin relaxation rate, $^{17}(1/T_1)_c$, and its c-axis Knight shift, $^{17}K_c$, are from Nandor et al~\cite{Nandor1999}. Since the $^{89}Y$ spin relaxation and Knight shift are relatively isotropic, the powder average spin relaxation, $^{89}(1/T_1)$, and shift, $^{89}K$, were also taken from Nandor et al~\cite{Nandor1999}. This O and Y data is from $100-700$ K.

The temperature dependence of the static spin magnetic susceptibility, $\chi_{\mathrm{spin}}$, between $300-700$ K, is discussed in Nandor et al and is important in leading to their conclusion that both MMP and ZBP fail. However, in the paper there is no plot or equation for it. In Valerie Nandor's PhD thesis~\cite{Nandorthesis} (see page 56), two equations are given for $\chi_{\mathrm{spin}}(T)$. One is extracted from the measured O Knight shift and the other is obtained from the measured Y Knight shift. These two expressions are very close to each other and their small difference may be considered an error estimate of $\chi_{\mathrm{spin}}(T)$.

The equation for $\chi_\mathrm{spin}(T)$ from $300-700$ K dervied from the $^{17}K_c$ data is

\begin{equation}
\label{chiO}
\frac{\chi_{\mathrm{spin}}}{\mu_B^2}=2.717(1-2.89\times 10^{-4}\mathrm{T})\ \mathrm{eV^{-1}}
\end{equation}

\noindent and the equation derived from the $^{89}K$ data is

\begin{equation}
\label{chiY}
\frac{\chi_{\mathrm{spin}}}{\mu_B^2}=2.705(1-2.75\times 10^{-4}\mathrm{T})\ \mathrm{eV^{-1}}.
\end{equation}

\noindent Here, $\mathrm{T}$ is the temperature in Kelvin and $\mu_B$ is the Bohr magneton. $\chi_{\mathrm{spin}}/\mu_B^2$ is the susceptibility in units of states per eV per planar Cu.

In this paper, we average equations \ref{chiO} and \ref{chiY} for fitting to $\chi_{\mathrm{spin}}$.

\section{Comparison of the Model to Experiment}

\subsection{NMR in the Normal State}

\begin{figure}[tbp]
\centering \includegraphics[width=1.0\linewidth]
{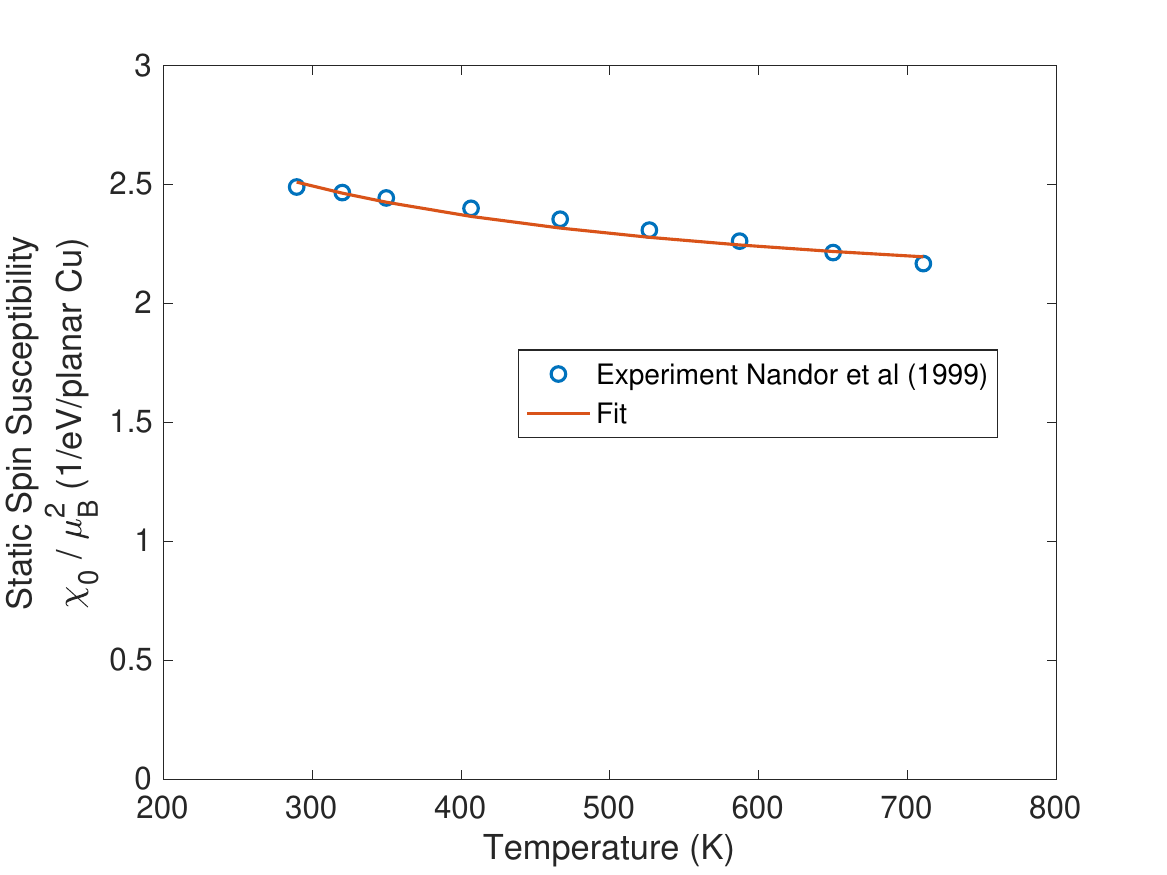}
\caption{Experimental static susceptibility between $300$ to $700$ K obtained by Nandor et al~\cite{Nandor1999,Nandorthesis} and its fit by equation~\ref{chif}. The experimental expression for $\chi(T)$ is the average of equations~\ref{chiO} and \ref{chiY} and is taken at the 9 temperature values (blue circles) used by Nandor et al for the O Knight shift shown in Figure~\ref{K17}. The fit is the red curve. The metallic density of states in the fit is $N(0)=1.512$ (1/eV/spin/planar metallic Cu site) and the dimensionles coefficient in $\chi_{AF}$ in equation~\ref{chiAF} is $\kappa=0.0482$. Tables~\ref{parshyp} and \ref{pars} list the final fitted parameters.
}
\label{chistatic}
\end{figure}

The static susceptibility expression in equation~\ref{chif} has three unknown parameters on the right-hand-side. They are the metallic density of the states, $N(0)$, appearing in $\chi_{M}$ in equation~\ref{chiM}, the energy scale for the inter-AF-cluster spin-spin coupling, $k_B T_x$, and the dimensionless constant, $\kappa$, that sets the magnitude of the AF static susceptibility in equation~\ref{chiAF} for $\chi_{AF}$.

\begin{figure}[tbp]
\centering \includegraphics[width=1.0\linewidth]
{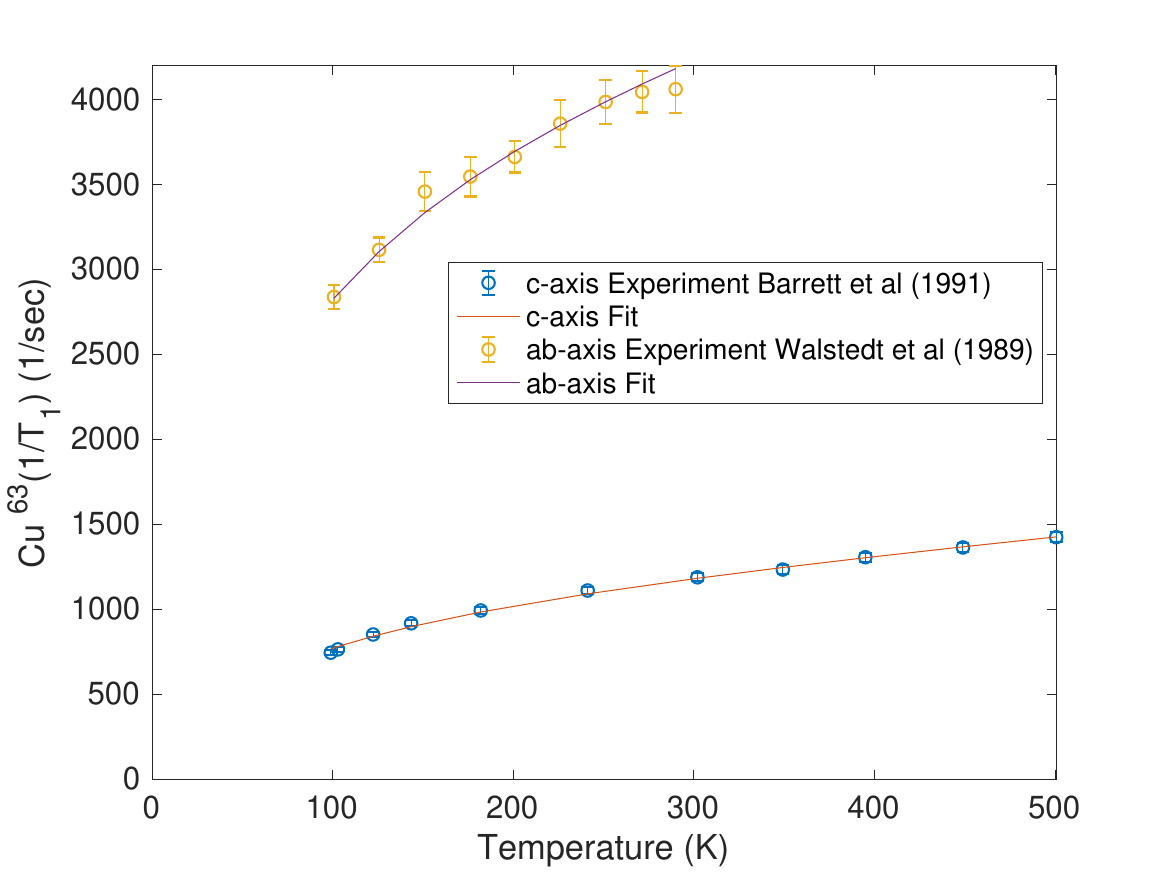}
\caption{Fit to the ab-axis and c-axis magnetic field $\mathrm{^{63}Cu}$ spin relaxation data. The c-axis data is taken from Barrett et al~\cite{Barrett1991} and the ab-axis data is from Walstedt et al~\cite{Walstedt1989}.  The metallic density of states was determined by the static susceptibility fit in Figure~\ref{chistatic} and was not adjusted to fit the Cu spin relaxations. The onsite hyperfine couplings, $A_c$ and $A_{ab}$, were fixed to the ZBP values and were not adjusted either. The metallic and AF transferred hyperfine couplings from the Cu $4s$ orbital were set equal to each other, $B_M = B_{AF}$, in the fit. Its fitted value is $B_M=B_{AF}=3.59\times 10^{-7}$ eV compared to the ZBP value of $B=4.0\times 10^{-7}$ eV. The temperature independent AF spin correlation length is $\xi_{AF}=2.05$ lattice spacings. The strength of the inter-AF-cluster spin coupling is $T_x=61$ K. The ``dwell time enhancement" factor due to coupling of the AF spins to the metallic region is $\lambda_{dwell}=2.5$. The high-temperature AF spin correlation time is $\tau_0=3.92\times 10^{-15}$ seconds. Tables~\ref{parshyp} and \ref{pars} list the final fitted parameters.
}
\label{Cu63}
\end{figure}

We fit the experimental static susceptibility by Nandor et al~\cite{Nandor1999,Nandorthesis} over $300-700$ K and the Cu spin relaxation rates with the magnetic field in the plane (ab-axis) and out of the plane (c-axis) and find $N(0)=1.512$ in units 1/eV/spin/planar metallic Cu, dimensionless $\kappa=0.0482$, and $T_x=61$ K. The fits are shown in figures~\ref{chistatic} and \ref{Cu63}.

In this fit, the onsite couplings $A_c$ and $A_{ab}$ were fixed to their ZBP values. The transferred hyperfine coupling was taken to be equal for the metallic and AF regions, or $B_{M}=B_{AF}\equiv B$. $B$ was allowed to vary from its ZBP value of $B_{ZBP}=4.0\times 10^{-7}$ eV and is fitted to $B=3.59\times 10^{-7}$ eV. We find the ``dwell enhancement factor" is $\lambda_{dwell}=2.5$. The high temperature AF spin correlation time is $\tau_0=3.92\times 10^{-15}$ seconds. The AF spin correlation length, $\xi_{AF}$, divided by the lattice spacing, $a$, is $\xi_{AF}/a=2.05$.

Earlier, we stated the order of magnitude of $\tau_0\sim\hbar/J_{AF}=5.06$ fs. Here, we estimate $\tau_0$ more carefully in order to compare it to our fitted value of $3.92$ fs. $\tau_0$ is given by,

\begin{equation}
\tau_0=\frac{1}{2}\int_{-\infty}^{+\infty}e^{-\frac{1}{2}\omega_e^2t^2}\ dt=\left(\frac{\pi}{2}\right)^{\frac{1}{2}}\left(\frac{1}{\omega_e}\right),
\label{tau0eqn}
\end{equation}

\noindent where $\omega_e$ is the spin correlation angular frequency from a high temperature Gaussian approximation to the spin autocorrelation function. The angular frequency is~\cite{Moriya1956a}

\begin{equation}
\omega_e^2=\left(\frac{J_{AF}}{\hbar}\right)^2\cdot 2z\cdot\left[\frac{1}{3}S(S+1)\right],
\label{omegae}
\end{equation}

\noindent where $z$ is the number of AF neighbors and $S=1/2$ is the electron spin. Note that Imai et al~\cite{Imai1993} are missing the factor of $2$ in their expression for $\omega_e^2$. Substituting equation~\ref{omegae} into equation~\ref{tau0eqn} leads to

\begin{equation}
\tau_0=\left(\frac{\pi}{z}\right)^{\frac{1}{2}}\left(\frac{\hbar}{J_{AF}}\right).
\label{tau0eqn1}
\end{equation}

The number of AF neighbors is $z=4p_{AF}=1.44$. Substituting into equation~\ref{tau0eqn1} leads to $\tau_0=7.48$ fs. This estimate of $\tau_0$ is almost two times as large as our fitted value. However, the Gaussian approximation~\cite{Moriya1956a} that led to equation~\ref{tau0eqn1} assumes $k_B T>>J_{AF}\approx 1500$ K. The NMR data is up to $700$ K. Hence, the Gaussian approximation to $\tau_0$ may be an upper bound.

Note that for a standard square AF lattice, $z=4$, leading to $\tau_0=(\pi^{1/2}/2)(\hbar/J_{AF})$. Monien et al~\cite{Monien1990} state in their equation 15 that $\tau_0=(\pi^{1/2}/4)(\hbar/J_{AF})$. This same expression appears in Pennington's thesis~\cite{Penningtonthesis} on page 75. We believe their equation is not correct. Pennington cites Horvati\'{c} et al~\cite{Horvatic1989} who cites Narath~\cite{Narath1967}. Monien et al also cite Narath. Narath in equation 53 on page 302 defines $J_{\mathrm{Narath}}$ to be the ``chemists" definition. Hence, $J_{AF}=2J_{\mathrm{Narath}}$. Narath merely states his result in equation 54 on page 302 with no derivation. Replacing the $J$ in Narath's expression with $(1/2)J_{AF}$ leads to our expression. We also derived equations~\ref{omegae} and ~\ref{tau0eqn1} ourselves in order to verify they are correct.

\begin{figure}[tbp]
\centering \includegraphics[width=1.0\linewidth]
{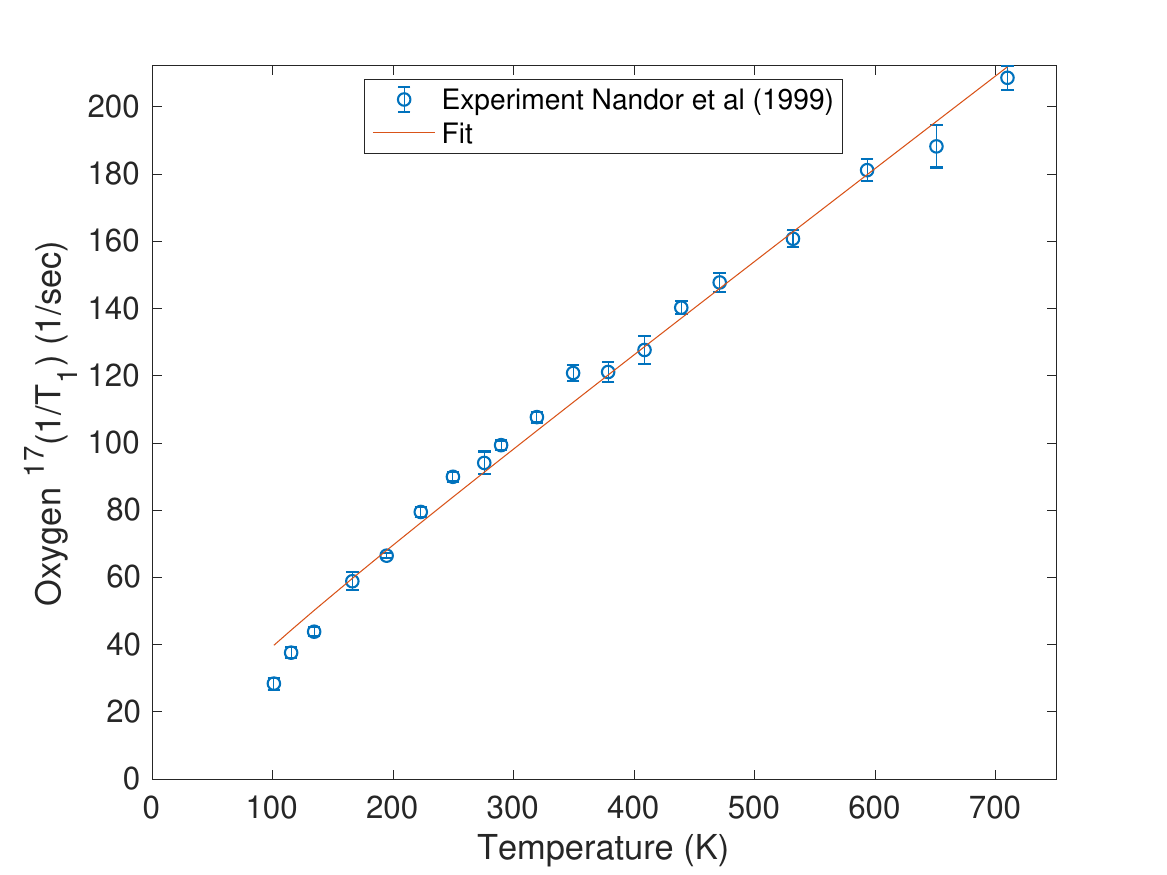}
\caption{Fit to the O c-axis spin relaxation data of Nandor et al~\cite{Nandor1999}. Two parameters affect this fit, $C_M$ and $C_{AF}$. The fit for $C_M=3.01\times 10^{-7}$ eV and $C_{AF}=2.08\times 10^{-7}$ eV is shown. As mentioned by Nandor et al, the calculated ZBP O spin relaxation rate is too small in magnitude by a factor of $\sim 2$. They state that this problem is not understood. Here, we fit both the spin relaxation rate and Knight shift without any corrections. Tables~\ref{parshyp} and \ref{pars} list the final fitted parameters.
}
\label{O17}
\end{figure}

\begin{figure}[tbp]
\centering \includegraphics[width=1.0\linewidth]
{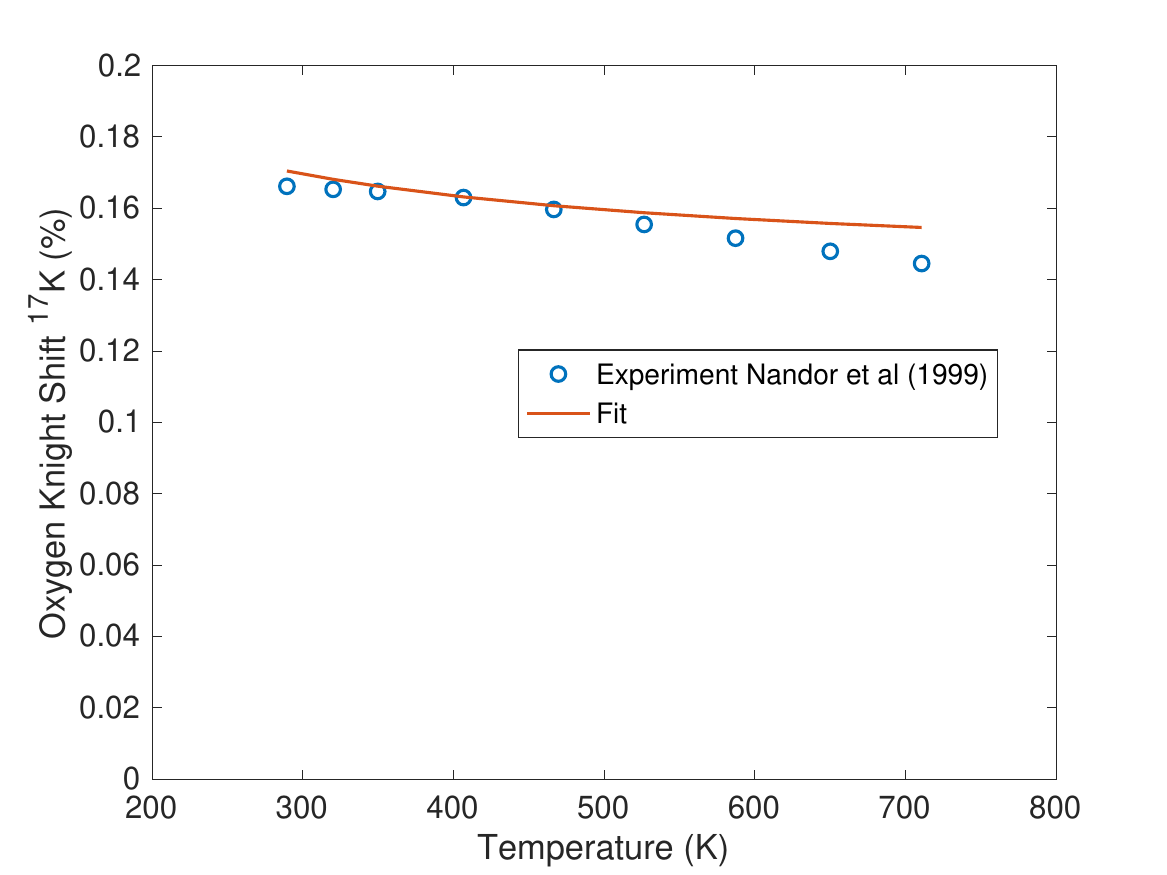}
\caption{Fit to the O Knight Shift data from Nandor et al~\cite{Nandor1999} from $300$ to $700$ K with $C_M=3.01\times 10^{-7}$ eV and $C_{AF}=2.08\times 10^{-7}$ eV. It is important to observe that the fits to the O Knight shift and its relaxation rate in Figure~\ref{O17} do not require an unexplained factor of $\sim 2$ correction, as discussed by Nandor et al~\cite{Nandor1999}, to make the relaxation and Knight shift magnitudes match experiment. Tables~\ref{parshyp} and \ref{pars} list the final fitted parameters.
}
\label{K17}
\end{figure}

\begin{figure}[tbp]
\centering \includegraphics[width=1.0\linewidth]
{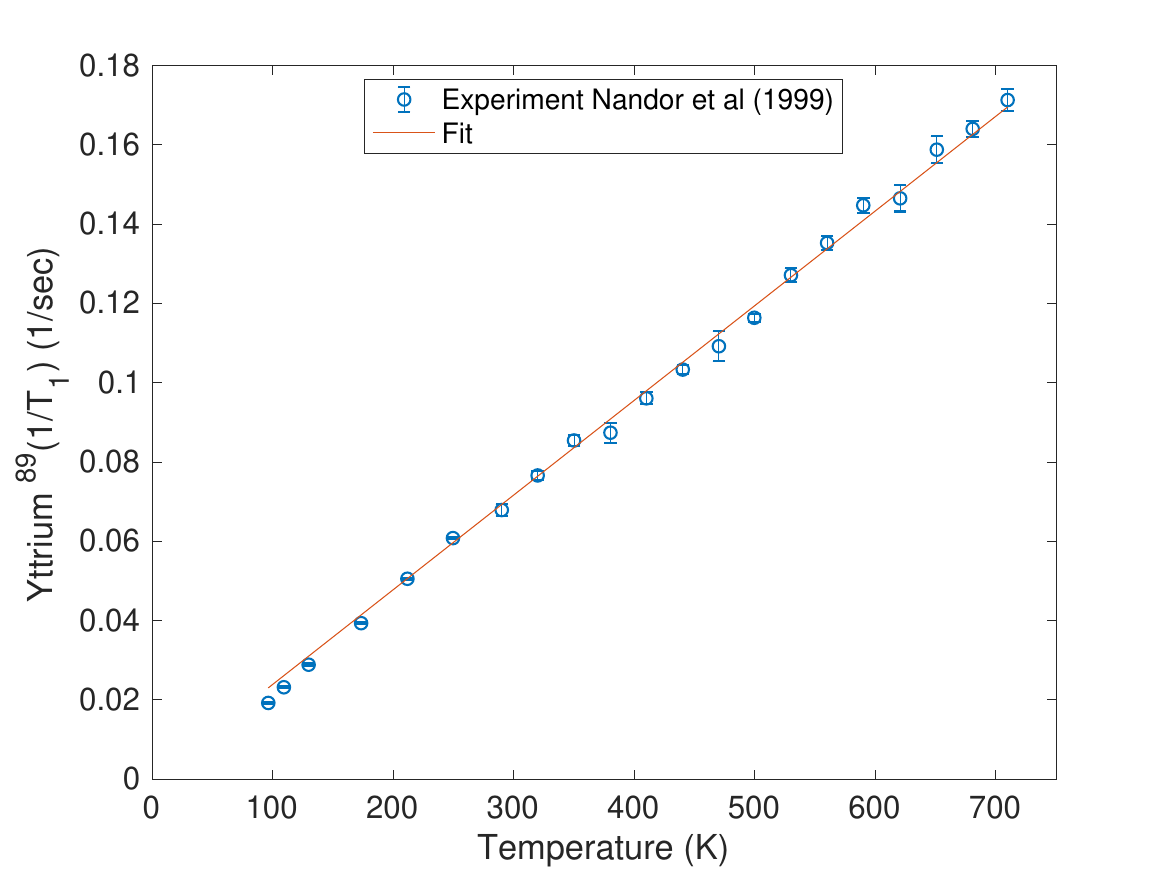}
\caption{Fit to the Y spin relaxation rate data assuming that the contribution from the AF spins is suppressed relative to the metallic electrons due to the use of the CPMG pulse sequence as discussed in the text. The fit is shown for $D_M=-4.45\times 10^{-9}$ eV. Tables~\ref{parshyp} and \ref{pars} list the final fitted parameters.
}
\label{Y89}
\end{figure}

\begin{figure}[tbp]
\centering \includegraphics[width=1.0\linewidth]
{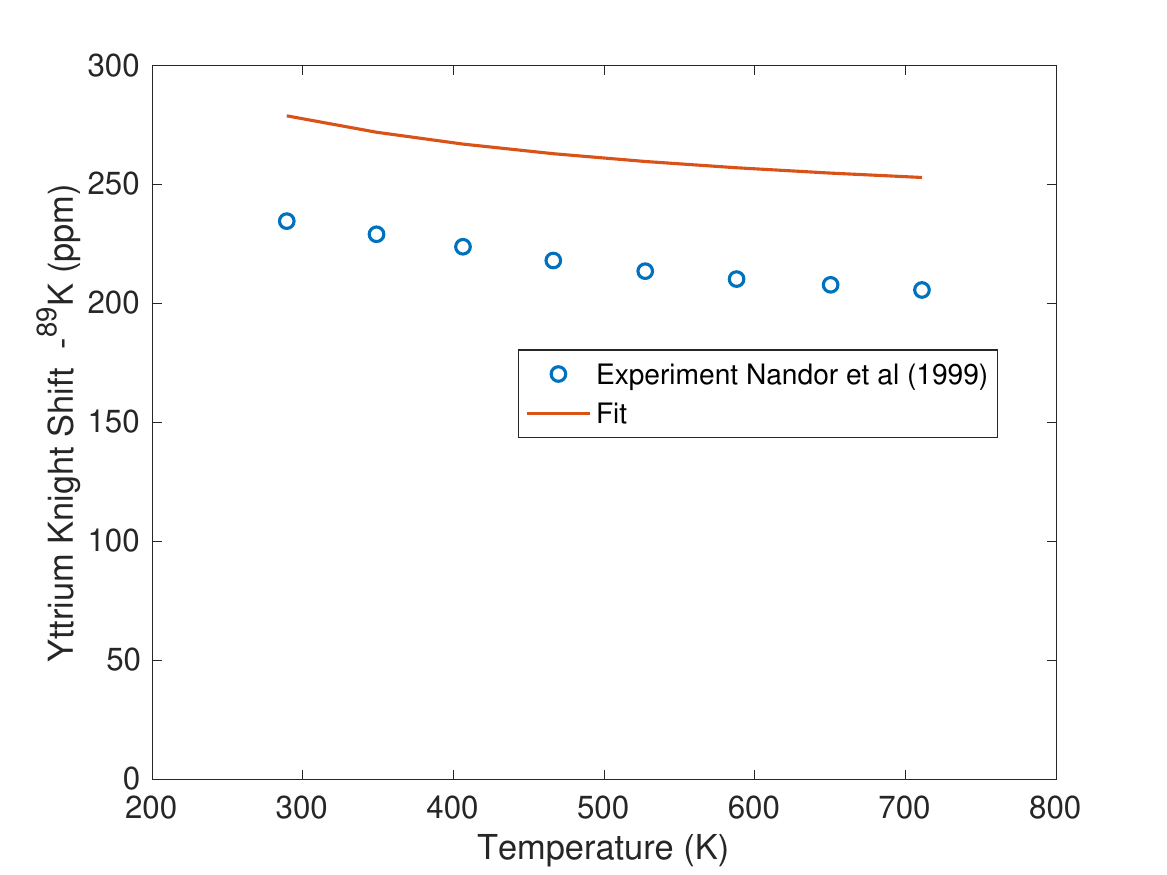}
\caption{Fit to Y shift data from Nandor et al~\cite{Nandor1999}. From equation~\ref{OYratio}, the final unknown parameter, $D_{AF}$ is determined to be $D_{AF}=-3.07\times 10^{-9}$ eV. Tables~\ref{parshyp} and \ref{pars} list the final fitted parameters.
}
\label{K89}
\end{figure}

To compare to the O spin relaxation and Knight shifts, there are only two remaining adjustable parameters. They are the magnitudes of the metallic, $C_M$, and AF region, $C_{AF}$, couplings to the O nucleus. With $C_M=3.01\times 10^{-7}$ eV and $C_{AF}=2.08\times 10^{-7}$ eV, the fit to the spin relaxation is shown in Figure~\ref{O17} and the fit to the Knight shift in Figure~\ref{K17}. The fit to the O spin relaxation rate is larger than experiment below $\approx 180$ K. We believe the experimental data drops faster than our fit due to a psuedogap that occurs in Nandor's $\mathrm{^{17}O}$ sample due to the sample having a slightly lower than optimal O concentration~\cite{Nandor1999}.

As discussed in the captions of figures~\ref{O17} and \ref{K17}, Nandor et al had to increase the ZBP predicted relaxation rate by a factor of $\sim 2$ in order to fit the data. No such correction is necessary for our fits. Due to the CPMG pulse sequence used by Nandor et al to obtain the Y spin relaxation up to $700$ K, we argued above that the contribution from the AF region is suppressed due to its shorter $T_2$ transverse relaxation time relative to the metallic $T_2$. Hence, only the metallic coupling to the Y nucleus, $D_M$, needs to be adjusted to fit the Y data. Figure~\ref{Y89} shows the fit with $D_M=-4.45\times 10^{-9}$ eV.

Since the same Cu orbitals lead to the O and Y couplings, we expect that the ratio of their metallic to AF coupling strengths should be the same. Hence we assume,

\begin{equation}
\label{OYratio}
\frac{D_{AF}}{D_M}=\frac{C_{AF}}{C_M}.
\end{equation}

Since $C_M$, $C_{AF}$, and $D_{M}$ are known, $D_{AF}$ is determined by equation~\ref{OYratio}. It leads to $D_{AF}=-3.07\times 10^{-9}$ eV. The fit to the Y Knight shift is shown in Figure~\ref{K89}.

Table~\ref{parshyp} compares the fitted hyperfine couplings to the ZBP parameters. Table~\ref{pars} shows the remaining 6 parameters necessary to define the dynamics of the electronic fluctuations in the metallic and AF regions.

\begin{table}[tbp]
\caption{Magnetic hyperfine couplings in units of eV. ZBP~\cite{ZBP} have three different couplings to the O atom, $\mathrm{C_c}$, $\mathrm{C_{||}}$, and $\mathrm{C_{\perp}}$ for the magnetic field along the c-axis, in the $\mathrm{CuO_2}$ plane parallel to the $\mathrm{Cu-O-Cu}$ bond direction, and in the $\mathrm{CuO_2}$ plane perpendicular to the $\mathrm{Cu-O-Cu}$ bond direction, respectively. The model in this paper sets all three of these parameters to be equal, $\mathrm{C_c=C_{||}=C_{\perp}\equiv C}$. This simplification is also used by MMP~\cite{MMP}. ZBP include a next-nearest Cu neighbor hyperfine coupling, $\mathrm{C'}$, to the O atom. In this paper, we set $\mathrm{C'=0}$.}
\label{parshyp}
\begin{tabular}{cc|c|c}
 &   & Fit & Fit \\
 & ZBP\footnote{The Zha-Barzykin-Pines (ZBP)~\cite{ZBP} parameters are from Nandor et al~\cite{Nandor1999}.} (eV) & AF Region (eV) & Metal Region (eV) \\
\hline
 $\mathrm{^{63}A_c}$\ \   & $\mathrm{-1.6\times 10^{-6}}$    & $\mathrm{-1.6\times 10^{-6}}$  & $\mathrm{-1.6\times 10^{-6}}$   \\
 $\mathrm{^{63}A_{ab}}$\ \  & $\mathrm{0.29\times 10^{-6}}$    & $\mathrm{0.29\times 10^{-6}}$  & $\mathrm{0.29\times 10^{-6}}$   \\
 $\mathrm{^{63}B}$\ \  &  $\mathrm{0.4\times 10^{-6}}$ & $\mathrm{0.359\times 10^{-6}}$  & $\mathrm{0.359\times 10^{-6}}$   \\
 $\mathrm{^{17}C_c}$\ \  & $\mathrm{0.156\times 10^{-6}}$    & $\mathrm{0.208\times 10^{-6}}$  & $\mathrm{0.301\times 10^{-6}}$   \\
 $\mathrm{^{17}C_{||}}$\ \  & $\mathrm{0.25\times 10^{-6}}$    & $\mathrm{0.208\times 10^{-6}}$  & $\mathrm{0.301\times 10^{-6}}$   \\
 $\mathrm{^{17}C_{\perp}}$\ \  & $\mathrm{0.13\times 10^{-6}}$    & $\mathrm{0.208\times 10^{-6}}$  & $\mathrm{0.301\times 10^{-6}}$   \\
 $\mathrm{^{89}D}$\ \  & $\mathrm{-4.8\times 10^{-9}}$    & $\mathrm{-3.07\times 10^{-9}}$  & $\mathrm{-4.45\times 10^{-9}}$   \\
\hline
\end{tabular}
\end{table}

\begin{table}[tbp]
\caption{Parameters that define the magnetic spin dynamics of the metallic and AF regions. The first column lists the parameter. The second column gives its value and dimensions. $\mathrm{\kappa}$, $\mathrm{\lambda_{dwell}}$, and $\xi_{AF}/a$ are dimensionless. The parameter, $a$, in the AF spin correlation length, $\xi_{AF}/a$, is the planar lattice spacing. The third column is the equation number that defines the parameter. The parameter, $\mathrm{N(0)}$, is the density of states per spin per energy per planar metallic Cu.}
\label{pars}
\begin{tabular}{ccc}
 Parameter\ \ \ &  Fitted Value\ \ \  & Defining Equation\ \ \ \\
\hline
 $\tau_0$ & $\mathrm{3.92\times 10^{-15}\ s}$ & \ref{tauaf} \\
 $T_x$    & 61 K & \ref{tauaf} \\
 $\kappa$ & 0.0482                    & \ref{chiAF} \\
 $\lambda_{dwell}$ & 2.5 & \ref{taum} \\
 $N(0)$   & $\mathrm{1.512\ eV^{-1}}$  & \ref{chiM} \\
 $\xi_{AF}/a$ & 2.05  & \ref{staticcorr} \\
\hline
\end{tabular}
\end{table}

There were two deep problems mentioned in the Introduction that led, in our opinion, to MMP and ZBP being abandoned. We already addressed the lack of tracking of $^{89}(1/T_1 T)$ and $\chi_{\mathrm{spin}}(T)$ by the different transverse relaxation rates, $T_2$, in the metal and AF regions and the use of the CPMG pulse sequence to obtain $^{89}(1/T_1)$.

The other problem was that the ZBP AF spin correlation length, $\xi\approx 2$ lattice spacings, was too large to be compatible with the neighboring $\mathrm{Cu-O}$ nuclear-nuclear coupling strength, $^{17,63}a$, relative to the $\mathrm{Cu-Cu}$ nuclear-nuclear spin coupling, $^{63,63}a$. Our fit to the Cu spin relaxation led to $\xi_{AF}=2.05$ lattice spacings. Hence, it would appear that the nuclear-nuclear coupling problem remains unexplained.

However, $\xi_{AF}$ applies only in the AF clusters. In the metallic clusters, the neighboring spins are weakly correlated. Similarly, the spin correlation between a metallic Cu atom neighboring an AF Cu atom is small. Therefore, the average spin correlation length is $\approx p_{AF}^2\xi_{AF}=(0.36)^2\cdot 2.05=0.266<<1$. Yu et al~\cite{Pennington-PRL-1999} and Pennington et al~\cite{Pennington-STIPDOR} concluded that a spin correlation length smaller than one lattice spacing was necessary to explain their data.

Before leaving the normal state NMR, we comment on the linear in $T$ contribution to the Cu spin relaxation rate (see equations~\ref{taum} and \ref{Am}) arising from metallic Cu neighbors. The linear $T$ slopes for the metallic Cu neighbors coupling to a metallic Cu atom nucleus are $0.97\ \mathrm{(sK)^{-1}}$ and $3.88\ \mathrm{(sK)^{-1}}$ for a c-axis and ab-axis magnetic field, respectively. For metallic Cu atoms coupling to an AF Cu nucleus, the value is independent of magnetic field direction and is $0.775\ \mathrm{(sK)^{-1}}$.

In a heroic analysis, Jurkutat et al~\cite{Jurkutat2019} looked at all of the normal state Cu spin relaxation data for both magnetic field directions. Their analysis ranged from slightly underdoped to overdoped superconducting phases. They also included data from overdoped $\mathrm{Tl_2Ba_2CuO_{6+y}}$ where $T_c=0$~\cite{Fujiwara1991}. They concluded that for all dopings, the cuprates must have a metallic (linear in $T$) contribution to the Cu spin relaxation rate that was independent of material and doping for the ab-axis $^{63}(1/T_1)_{ab}$ with slope $\approx 21\ \mathrm{(sK)^{-1}}$. Their analysis and conclusion are very compelling.

Our values of $3.88\ \mathrm{(sK)^{-1}}$ and $0.775\ \mathrm{(sK)^{-1}}$ quoted above are not close to Jurkutat et al's value. In fact, our small slopes are compatible with the observation of a linear $T$ slope up to $700$ K, as discussed by Nandor et al~\cite{Nandor1999} in their Figure 9.

We were unable to explain the NMR relaxation and shift data up to temperatures far above room temperature ($500$ K for Cu and $700$ K for O and Y) assuming the large slope of Jurkutat et al. In our model, the overdoped $\mathrm{Tl_2Ba_2CuO_{6+y}}$ with $T_c=0$ is comprised of mostly metallic Cu atoms and very few AF Cu atoms. In order to obtain the observed relaxation slope, the amount of Cu $4s$ character must have increased from its fraction at optimally doped $\mathrm{YBa_2Cu_3O_7}$.

Mackenzie et al~\cite{Mackenzie1995,Mackenzie1996} found that Cu substitutes at the Tl site in Tl2201 with $\approx 5.5\%$ substitution for the tetragonal phase and $<1\%$ for the orthorhombic phase. Hence, it is unclear if the observed planar Cu spin relaxation (there is no chain Cu in this material) is due to the planar Cu atom or the Cu atom at the Tl site. The latter Cu atom may have more $4s$ character at the Fermi level leading to the large slope quoted by Jurkuatat et al~\cite{Jurkutat2019}.

A further weakness with Jurkutat et al's picture is that it does not explain why the metallic linear slope substantially decreases at higher temperatures. This slope decrease suggests that their attribution of the full ab-axis relaxation rate magnitude just above $T_c$ to a ``metallic" effect may not be correct.

Finally, our small slopes are more compatible with the small slopes found in NQR for optimally doped $\mathrm{La_{2-x}Sr_{x}CuO_4}$~\cite{Imai1993} and $\mathrm{YBa_2Cu_4O_8}$~\cite{Curro1997,Tomeno1994}. In the latter case the material is underdoped. Hence, it has a smaller fraction of metallic sites leading to a smaller metallic slope.

\subsection{Spin Relaxation in the Superconducting State}

Below $T_c$, a d-wave superconducting gap appears in the metallic region. Since the AF clusters do not percolate in a $\mathrm{CuO_2}$ plane, any ``inter-AF-cluster" spin-spin coupling is mediated by the available electrons in the metallic region. This coupling is proportional to the metallic susceptibility near $\mathbf{k}$ vectors close $(\pm\pi/a,\pm\pi/a)$ since the AF wave-vector is peaked there. The opening of a d-wave superconducting gap strongly suppresses the metal susceptibility near $(\pm\pi/a,\pm\pi/a)$ leading to a decoupling of the AF clusters with each other with decreasing $T$. Hence, each AF cluster disconnects from the metallic region and the other AF clusters at low temperatures.

The spin eigenstates in a finite AF cluster are not a continuum of energies. Instead, the eigenstate energies are discrete with an energy separation that is orders of magnitude larger than the nuclear energy splittings. A spin gap emerges between the ground spin state in each cluster and its first excited state as the AF clusters decouple from neighboring AF clusters for $T<T_c$. The opening of a spin gap in the superconducting state has been seen by neutron scattering on optimally doped $\mathrm{La_{2-x}Sr_xCuO_4}$ with $x=0.16$~\cite{Christensen2004} and close to optimally doped $\mathrm{YBa_2Cu_3O_{6.85}}$~\cite{Bourges2000}. Therefore, the spin relaxation rate of the nuclei in the AF clusters goes to zero rapidly with the opening of a superconducting gap.

The $\mathrm{^{63}Cu}$ spin relaxation rate anistropy is,

\begin{equation}
\label{anisotropy}
\frac{^{63}(1/T_1)_{ab}}{^{63}(1/T_1)_c}= \frac{p_M{^{63}(1/T_1)_{ab,M}} + p_{AF}{^{63}(1/T_1)_{ab,AF}}}{p_M{^{63}(1/T_1)_{c,M}} + p_{AF}{^{63}(1/T_1)_{c,AF}}},
\end{equation}

\noindent where $p_M=0.64$ and $p_{AF}=0.36$. From our fits to the normal state NMR data, at $T\approx 100$ K, $^{63}(1/T_1)_{c,M}=875$ $\mathrm{s^{-1}}$, $^{63}(1/T_1)_{c,AF}=575$ $\mathrm{s^{-1}}$, $^{63}(1/T_1)_{ab,M}=1175$ $\mathrm{s^{-1}}$, and $^{63}(1/T_1)_{ab,AF}=5774$ $\mathrm{s^{-1}}$.

For a d-wave gap, the $T$ dependence below $T_c$ of the metallic spin relaxation rates scales with $T$ approximately as $^{63}(1/T_1)_{M}\sim T^3$. The AF spin relaxation rates goes to zero rapidly and is likely of the form $^{63}(1/T_1)_{AF}\sim e^{-\beta\Delta}$ where $\beta=1/k_B T$ and $\Delta$ is the magnitude of the d-wave gap. The AF relaxation rate ratio drops below $T_c$ because the $(\pm\pi/a,\pm\pi/a)$ scatterings across the Fermi surface are suppressed as a d-wave gap opens and this suppression reduces the ab-axis relaxation rate faster than the c-axis relaxation rate for the AF nuclear spins. This explanation is identical to references~\cite{Bulut1992} and \cite{Thelen1993}. Since the ab-axis AF spin relaxation dominates over the metallic relaxation rates and the c-axis AF relaxation rate, the total spin relaxation anisotropy will decrease rapidly as $T$ decreases below $T_c$.

At very low temperature, the AF spin relaxation is negligible and the spin relaxation rate anisotropy becomes,

\begin{equation}
\label{anisotropym}
\frac{^{63}(1/T_1)_{ab}}{^{63}(1/T_1)_c}\approx\frac{{^{63}(1/T_1)_{ab,M}}}{^{63}(1/T_1)_{c,M}},
\end{equation}

Near $T=0$, the only electronic states available for flipping the Cu nuclear spin are along the $\mathbf{k}$-vector diagonals ($k_x=\pm k_y$). Along the diagonals, there is no Cu $4s$ character because a $d_{x^2-y^2}$ and $4s$ transform differently under diagonal reflection (interchange of $x$ and $y$, $x\leftrightarrow y$). Thus, the band eigenstates along the diagonals are purely Cu $d_{x^2-y^2}$ with no Cu $4s$ character.

However, the metallic region does not have perfect translational symmetry. Hence, momentum $\mathbf{k}$ is not a perfect quantum number. It is an approximate quantum number that leads to a minimum metallic Cu $4s$ hyperfine coupling, $B_{M,\mathrm{min}}$, for band states near the diagonals. Hence, the spin relaxation anisotropy at very low temperatures is,

\begin{equation}
\label{anisotropym1}
\frac{^{63}(1/T_1)_{ab}}{^{63}(1/T_1)_c}=\frac{\left(A_c^2+n_M B_{M,\mathrm{min}}^2\right)+\left(A_{ab}^2+n_M B_{M,\mathrm{min}}^2\right)}{2\left(A_{ab}^2+n_M B_{M,\mathrm{min}}^2\right)},
\end{equation}

\noindent where $n_M=4p_M=2.56$ is the average number of metallic Cu neighbors to the nucleus.

$B_{M,\mathrm{min}}$ should be between zero and $B_{M}$, or $0<B_{M,\mathrm{min}}/B_M < 1$. For $B_{\mathrm{min}}=0$, the spin anisotropy ratio in equation~\ref{anisotropym1} is $15.72$, and when $B_{\mathrm{min}}=B_M$, the ratio is $3.99$. Therefore, the Cu spin relaxation anisotropy should converge to a value between $3.99$ and $15.72$ as $T\rightarrow 0$.

The above analysis leads to the following scenario for the temperature dependence of the Cu spin relaxation anisotropy ratio below $T_c$. Just above $T_c$, the ratio is $\approx 3.7$. As $T$ decreases below $T_c$, the opening of a d-wave supercoducting gap in the metallic region leads to a rapid exponential drop of the AF contributions to the relaxation because the inter-AF-cluster spin-spin coupling is mediated by the metallic electrons and there are fewer of them at the Fermi energy as the gap increases in magnitude. The metallic region's contribution to the anisotropy changes more slowly as a power law $\sim T^3$. Hence, the anisotropy will initially decrease as $T$ decreases just below $T_c$.

As $T$ continues to decrease, the AF contribution to the spin relaxation goes to zero, and the ratio is dominated by the metallic region. This ratio will converge to a value between $3.99$ and $15.72$ as $T\rightarrow 0$. Hence, the spin relaxation anisotropy ratio will rise back up as $T$ decreases and eventually converge to a value larger than the normal state ratio of $\approx 3.7$.

Experiments on $\mathrm{YBa_2Cu_3O_7}$~\cite{Slichter1993,Martindale1991,Takigawa1991} and intrinsically underdoped $\mathrm{YBa_2Cu_4O_8}$~\cite{Brinkmann1992} match this picture. For $\mathrm{YBa_2Cu_3O_7}$ the $T=0$ ratio is $\approx 5$ and for $\mathrm{YBa_2Cu_4O_8}$ the ratio is $\approx 6$. Since $\mathrm{YBa_2Cu_4O_8}$ is underdoped, the number of metallic neighbors will be smaller than its value for $\mathrm{YBa_2Cu_3O_7}$, or $n_{M}<2.56$. A smaller value of $n_M$ in equation~\ref{anisotropym1} leads to a larger anisotropy ratio assuming $B_{\mathrm{min}}$ is unchanged.

\section{Conclusions}

The idea that there are two electronic components in cuprates is not original to this paper. In the past decade, Haase and collaborators have made many strong arguments for two spin components in cuprates. What is different here is that we propose that the two electronic spin components create atomic-scale inhomogeneity that provides the additional degrees of freedom necessary to understand the cuprate NMR.

In this paper, we have provided a specific physical model of the two components and the nature of their atomic-scale inhomogeneity. Using this picture, we have shown that a broad spectrum of cuprate NMR phenomenology can be understood. This paper focused on the cuprate NMR up to the early 2000s. The astonishing body of work by Haase and collaborators in the last $\sim 15$ years has not yet been analyzed.

\section*{Acknowledgments}

It is shocking to me that the NMR data on cuprates is ignored by cuprate theorists. Questioning the wisdom of the cuprate academic Ancien R\'{e}gime is above my pay grade. I am just thankful for the resulting opportunity.

I am extremely grateful to Michael Jurkutat for several long discussions about the cuprate NMR over the past few years. His knowledge and willingness to share his time were invaluable. Thanks also to Sean E. Barrett for discussing his superconducting state NMR data from $\sim 1990$ with me. A special thanks to Carver A. Mead for discussions on all aspects of the NMR problem and also for his financial support. His advice, ``Better to be a rebel than some rubble!"~\cite{CAM}, is very appropriate here.

\bibliography{nmr.bbl}

\end{document}